\documentclass[sigconf]{acmart}

\usepackage{url}
\usepackage{array,framed}
\usepackage{enumitem, kantlipsum}

\usepackage{
  color,
  float,
  epsfig,
  wrapfig,
  graphics,
  graphicx,
  subcaption
}

\usepackage{latexsym,fancyhdr}
\usepackage{algpseudocode}
\usepackage{multirow}
\usepackage{hyperref}

\usepackage{
  tikz,
  pgfplots,
  pgfplotstable
}

\usetikzlibrary{
  shapes.geometric,
  arrows,
  external,
  pgfplots.groupplots,
  matrix
}

\pgfplotsset{compat=1.9}

\usepackage{mathtools}

\DeclareMathAlphabet{\mathcal}{OMS}{cmsy}{m}{n}

\DeclareGraphicsExtensions{%
    .png,.PNG,%
    .pdf,.PDF,%
    .jpg,.mps,.jpeg,.jbig2,.jb2,.JPG,.JPEG,.JBIG2,.JB2}

\usepackage{xparse}
\newcommand{\bnm}{\begin{newmath}}
\newcommand{\enm}{\end{newmath}}

\newcommand{\bea}{\begin{eqnarray*}}%
\newcommand{\eea}{\end{eqnarray*}}%

\newcommand{\bne}{\begin{newequation}}
\newcommand{\ene}{\end{newequation}}

\newcommand{\bal}{\begin{newalign}}
\newcommand{\eal}{\end{newalign}}

\newenvironment{newalign}{\begin{align}%
\setlength{\abovedisplayskip}{4pt}%
\setlength{\belowdisplayskip}{4pt}%
\setlength{\abovedisplayshortskip}{6pt}%
\setlength{\belowdisplayshortskip}{6pt} }{\end{align}}

\newenvironment{newmath}{\begin{displaymath}%
\setlength{\abovedisplayskip}{4pt}%
\setlength{\belowdisplayskip}{4pt}%
\setlength{\abovedisplayshortskip}{6pt}%
\setlength{\belowdisplayshortskip}{6pt} }{\end{displaymath}}

\newenvironment{newequation}{\begin{equation}%
\setlength{\abovedisplayskip}{4pt}%
\setlength{\belowdisplayskip}{4pt}%
\setlength{\abovedisplayshortskip}{6pt}%
\setlength{\belowdisplayshortskip}{6pt} }{\end{equation}}

\newcounter{ctr}

\newcounter{mytable}
\def\mytable{\begin{centering}\refstepcounter{mytable}}
\def\endmytable{\end{centering}}

\newcounter{myfig}
\def\myfig{\begin{centering}\refstepcounter{myfig}}
\def\endmyfig{\end{centering}}

\newlength{\saveparindent}
\newlength{\saveparskip}
\newcommand{\E}{{\rm I\kern-.3em E}}

\renewcommand{\eqref}[1]{\mbox{Equation~(\ref{#1})}}

\def \part {part}

\renewcommand{\paragraph}[1]{\vspace*{6pt}\noindent\textbf{#1}\;}

\def \blackslug{\hbox{\hskip 1pt \vrule width 4pt height 8pt
    depth 1.5pt \hskip 1pt}}
\def \qed{\quad\blackslug\lower 8.5pt\null\par}
\newcounter{mynote}[section]

\newcommand\ignore[1]{}

\newcounter{rcnote}[section]

\newcounter{mrnote}[section]

\newcounter{fknote}[section]

\newcounter{anote}[section]

\DeclareMathSymbol{\mlq}{\mathord}{operators}{``}
\DeclareMathSymbol{\mrq}{\mathord}{operators}{`'}

\newcommand{\rhf}[2]{R_{f, \gamma}}

\newcommand{\M}{{\mathcal{M}}}

\newcommand{\m}{{w}}

\DeclareDocumentCommand{\edist}{o o}{
  \ensuremath{
    \IfNoValueTF{#1}{{d}}{{\sf d}(#1,#2)}
  }
}

\newcommand{\s}{\mathcmd{s}}

\newcommand{\olrk}[1]{\ifx\nursymbol#1\else\!\!\mskip4.5mu plus 0.5mu\left(\mskip0.5mu plus0.5mu #1\mskip1.5mu plus0.5mu \right)\fi}

\NewDocumentCommand{\indseq}{ O{1} O{r} }{{#1}\ldots {#2}}

\usepackage{pifont}

\newcommand{\subhead}[1]{\vspace{0.5mm} \noindent{\textbf{#1.}}}

\def\S{Super-app}
\def\s{super-app}
\def\M{Mini-app}
\def\m{mini-app}
\def\a{app secret}

\copyrightyear{2023}
\acmYear{2023}
\setcopyright{acmlicensed}\acmConference[RAID '23]{The 26th International
Symposium on Research in Attacks, Intrusions and Defenses}{October 16--18,
2023}{Hong Kong, Hong Kong}
\acmBooktitle{The 26th International Symposium on Research in Attacks,
Intrusions and Defenses (RAID '23), October 16--18, 2023, Hong Kong, Hong
Kong}
\acmPrice{15.00}
\acmDOI{10.1145/3607199.3607236}
\acmISBN{979-8-4007-0765-0/23/10}

\begin{document}

\author{Supraja Baskaran}
\affiliation{%
  \institution{Concordia University}
  \city{Montreal}
  \country{Canada}}
   \email{{su_baska@ciise.concordia.ca}}
\author{Lianying Zhao}
\affiliation{%
  \institution{Carlton University}
  \city{Ottawa}
  \country{Canada}}
   \email{{lianying.zhao@carleton.ca}}
\author{Mohammad Mannan}
\affiliation{%
  \institution{Concordia University}
  \city{Montreal}
  \country{Canada}}
   \email{{mmannan@ciise.concordia.ca}}
\author{Amr Youssef}
\affiliation{%
  \institution{Concordia University}
  \city{Montreal}
  \country{Canada}}
   \email{{youssef@ciise.concordia.ca}}

\title[Measuring the Leakage and Exploitability of Authentication Secrets in \S{}s]{Measuring the Leakage and Exploitability of Authentication Secrets in \S{}s: The WeChat Case}

\begin{abstract}
\S{}s such as WeChat and Baidu host millions of \m{}s, which are very popular among users and developers because of the \m{}s' convenience, lightweight, ease of sharing, and not requiring explicit installation. Such ecosystems involve several entities, such as the \s{} and \m{} clients, the \s{} backend server, the \m{} developer server, and other hosting platforms and services used by the \m{} developer. To support various user-level functionalities, these components must authenticate each other, which differs from regular user authentication to the \s{} platform. 
In this paper, we explore the \m{} to super-app authentication problem caused by insecure development practices. This type of authentication allows the \m{} code to access super-app services on the developer's behalf.

We conduct a large-scale measurement of developers' insecure practices leading to \m{} to \s{} authentication bypass, among which hard-coding developer secrets for such authentication is a major contributor. We also analyze the exploitability and security consequences of developer secret leakage in \m{}s by examining individual \s{} server-side APIs. We develop an analysis framework for measuring such secret leakage, and primarily analyze 110,993 WeChat \m{}s, 
and 10,000 Baidu \m{}s (two of the most prominent \s{} platforms), along with a few more datasets to test the evolution of developer practices and platform security enforcement over time. We found a large number of WeChat \m{}s (36,425, 32.8\%) 
and a few Baidu \m{}s (112) leak their developer secrets, which can cause severe security and privacy problems for the users and developers of \m{}s. A network attacker who does not even have an account on the \s{} platform, can effectively take down a \m{}, send malicious and phishing links to users, and access sensitive information of the \m{} developer and its users. 
We responsibly disclosed our findings and also put forward potential directions that could be considered to alleviate/eliminate the root causes of developers hard-coding the \a{}s in the \m{}'s front-end code.

\end{abstract}

\begin{CCSXML}
<ccs2012>
   <concept>
       <concept_id>10002978.10003022</concept_id>
       <concept_desc>Security and privacy~Software and application security</concept_desc>
       <concept_significance>500</concept_significance>
       </concept>
   <concept>
       <concept_id>10002978.10002991.10002992</concept_id>
       <concept_desc>Security and privacy~Authentication</concept_desc>
       <concept_significance>500</concept_significance>
       </concept>
 </ccs2012>
\end{CCSXML}

\ccsdesc[500]{Security and privacy~Software and application security}
\ccsdesc[500]{Security and privacy~Authentication}

\keywords{Authentication, Mini-app Security, WeChat, Hard-coded Secrets}
\maketitle

\section{Introduction}
\label{sec:introduction}

Full-featured apps such as WeChat \cite{wechatintro} and  Baidu~\cite{baiduintro}, 
with a monthly user base of over one billion~\cite{wechatpopularity,baidupopularity}, have created an ecosystem to accommodate payments, media, online stores, developers, etc~\cite{lien2014examining, guo2018extraversion, zheng2019mega}. Such popularity and self-contained ecosystem have enabled them to become ``\s{}s'', serving as the hosting platform of millions of \m{}s (also known as mini-programs). \M{}s, together with their \s{}s, are seeing increasing demands in different countries because of their convenience, light weight, ease of sharing and no need to install~\cite{minicrawler}.
In addition, \m{}s do not need a custom backend server from developers, but use a well-constructed set of APIs provided by their \s{} to allow straightforward access to backend data and system resources that are provided by the \s{} platforms.

Unsurprisingly, security and privacy issues also start to surface in these \s{} platforms, 
e.g., users' PII (name, national ID, date of birth, and facial data) collection by WeChat \m{}s~\cite{privacymini}.
More extensive security-focused work in this domain includes (details in Sec.~\ref{sec:relwork}): analysis of \m{} permission models~\cite{zhang2022small}, identity confusion attacks~\cite{zhang2022identity}, and 
cross \m{} request forgery attacks~\cite{yang2022cross}.

In contrast to existing work, we focus on the \m{}'s authentication secret (developer secret or \a{}) leakage, i.e., the secrets used to authenticate any part of the \m{} code to the \s{} server that it is the \m{} it claims to be. This is complicated by several factors, including: 
lack of a human user presenting secrets at the time of authentication (hence the need for somewhere to store the \m{} to \s{} authentication secret),
\m{} developers not following security guidelines from \s{} platforms, failure of \s{} platforms to enforce necessary security restrictions, multiple entities being involved in a \s{} ecosystem (e.g., \m{} client, cloud functions, server-side APIs, and data storage provided by a \s{} server, \m{} developer server), which also authenticate each other--either explicitly or implicitly.

A common way for the \s{} to authenticate a \m{} is to use \a{}s.
According to the WeChat documentation~\cite{securitydoc}, 
an \a{} is specific to a developer account, and it is used to authenticate any part of the \m{} (representing the developer) to the \s{}'s server. Therefore, the \a{} should be treated as a sensitive piece of data and should not be exposed (e.g., in the \m{} source code). 
Due to the absence of a human user in such authentication scenarios, \a{}s have to be stored instead of being entered by a user. 
WeChat expects \m{}s to offload such a burden to the developer's server which stores \a{}s and performs necessary communication with the \s{} server, i.e., not to hard-code in the \m{} code. 
For example, in the case of WeChat \m{}s, the \m{} packages can be extracted from a rooted/jail-broken device, or using the PC client, with common reverse-engineering techniques without any special privilege~\cite{frida-old} 
and the \a{}s, if hard-coded, can be easily obtained by anyone. %
If the developers still use such hard-coding,
since the \s{} server has complete control over the \s{} ecosystem, it can easily spot such hard-coding and prevent it either before releasing a \m{} or at run-time.
Our work is motivated by the observation that it is not the case in reality. %

When such \m{} to \s{} authentication is compromised due to hard-coded \a{}s, one can intuitively imagine how things may go wrong afterwards, e.g., the attacker will be able to impersonate the legitimate \m{} (developer) and manipulate/abuse its resources (e.g., images or order info). %
To better understand this problem, we examine how \a{}s are used to achieve the authentication:
it is the \s{} server that authenticates a remote party claiming to be the \m{} and provides subsequent services all through a set of exposed APIs over the network~\cite{api-class}. Therefore, these APIs eventually become the target of the authentication compromise, i.e., whether they can be invoked in an unauthorized manner.
The \s{} platforms often have certain security guidelines to ensure proper \m{} to \s{} authentication. For instance, WeChat discourages having \a{}s in the \m{}, recommends the use of \emph{IP address whitelisting} for such APIs (not allowing calls from non-listed addresses, configured at the developer portal), 
and disallows directly calling such APIs from within the \m{} (as opposed to doing it from the developer's server)~\cite{securitydoc}.
Still, enforcement of these guidelines remains a question, leading to insecure development practices.

The \m{} to \s{} authentication mechanisms vary across \s{} platforms, and thus the ways of managing authentication secrets vary. For example, Douyin~\cite{duoyinintro} (the Chinese version of Tiktok~\cite{tiktokintro}), uses a similar authentication technique like WeChat and Baidu. The \m{} to \s{} authentication system, which relies on \texttt{"client\_credential"} grant type to generate access tokens, is significantly affected by the \m{} developers' mishandling of the secret. 
In contrast, Alipay employs a dynamic authorization token for the generation of access token which involves no static secrets (\texttt{"authorization\_code"} grant type), thus providing a better authentication with the \s{}. 
Our analysis targets major popular \s{} ecosystems that have such authentication secret leakage problems, although primarily focuses on WeChat as it is the largest among these platforms in terms of \m{} count~\cite{wechatpopularity}.

Our objective is to measure the extent to which the aforementioned insecure \m{} development practices are found along the timeline of recent years (assuming the awareness is improving), and how such practices could have led to potential unauthorized calls to the \s{} server APIs. We also explore the security consequences of such calls in the \s{} ecosystem, e.g., business/personal resources in various application scenarios.
To do so, we develop an analysis framework that automatically detects \a{}s in the \m{}'s code through static analysis and verifies if authentication bypass is possible to call unauthorized \s{} server APIs, confirming the validity of the identified secrets and the lack of IP whitelisting. %
For ethical reasons, we do not make calls to all the \s{} server-side APIs, but instead, divide them into Get (can only read data) and Modify (can update/delete data) APIs. 
The framework automatically makes calls to only the necessary Get APIs, which provides  the required parameters to perform a ``callability''  analysis of Modify APIs. 
Also, we analyze and categorize the security consequences of the unauthorized \s{} server-side API access and shed light on the ways forward for improvement.

\vspace{10pt}
\subhead{Summary of contributions and notable findings}
\begin{enumerate}[wide, labelwidth=!, labelindent=7pt]
\vspace{-5pt}
\item We examine the mishandling of authentication secrets in WeChat and Baidu \m{}s, two leading \s{} providers.
The identified issue eventually leads to unauthorized \s{} server-side API calls by any network attacker, allowing access to various \m{}-owned resources.
Our methodology is designed to facilitate \emph{automated} analysis of \m{}s
from \s{} platforms that have a similar \m{} directory structure (like WeChat~\cite{weixindirstructure} and Baidu~\cite{baidudirstructure}), and use secret-based \m{} authentication. 

\item %
We conduct a large-scale automated measurement to assess the extent of these insecure practices in WeChat and Baidu. %
Out of 110,993 WeChat \m{}s that we could successfully decrypt and unpack (from a total of 115,392, crawled in 2021) 
and 10,000 Baidu \m{}s, we found a large number of WeChat \m{}s (36,425, 32.8\%) 
and a few Baidu \m{}s (112) leak their developer secrets, which can cause authentication bypass. The use of IP whitelisting, a WeChat security feature, which could restrict exploitation of such secret exposures, is also very limited--only 33 out of 110,993 \m{}s have enabled it (7 \m{}s with \a{}). We also automatically check data leakage through the available Get APIs, and callability of Modify APIs that can directly interfere with a \m{} functionality. We test the effects of dangerous Modify APIs only on our own \m{}.

\item From our responsible disclosure, we learned that WeChat is aware of \a{} hard-coding (independent of our reporting), and has enforced a new requirement that publishing such \m{}s are disallowed (which we also confirmed by submitting a new \m{} with hard-coded \a{}--it was rejected). To check the effectiveness of this new requirement, and to see how developer behaviours have evolved over the years, we performed a few additional measurements. From 9,824 \m{}s crawled in Feb.\ 2023, 2,572 (26.2\%) have valid \a{}s, and from the 36,425 \m{}s (crawled in 2021, with valid \a{}s), 36,293 (99.6\%) still have valid \a{}s---making recent enforcement by WeChat largely ineffective against existing vulnerable \m{}s (which were made public before Mar.\ 2023).

\item We conduct an in-depth attack feasibility analysis for individual APIs and \m{}s, and categorize the security consequences of such attacks. The consequences vary a lot with the semantics of the involved APIs, the configuration and functions of individual \m{}s, and several types of consequences are high-impact and affect a large number of \m{}s. 

\item We discuss the root causes of the hard-coding \a{}s, and suggest several recommendations for design and enforcement considerations. We will make our tool available to any \s{} platforms, who can identify and measure security consequences for different apps and take appropriate mitigating actions.  

\end{enumerate}

\section{Background} \label{sec:background}
We first briefly explain the various terms/entities involved in \m{} ecosystems, specifically in WeChat and Baidu.

\subhead{\S{}} %
A \emph{\s{} client} is the host mobile app that features a selection of independent services, all contained within a single app.
The \emph{\s{} server} is the key
authority of the \s{} platform, managing identity and operations of the \m{}s, and providing necessary services.
The \s{} server exposes a standard set of APIs to all \m{}s, which we refer to as \emph{server-side APIs} hereafter.

\subhead{\M{}} \label{miniapp-client}
The \emph{\m{} client} is the client-side code that runs on top of the \s{} client (also called mini-program, smart-program, micro-app). It is created with the corresponding \s{}'s devtools and shipped as a package (JavaScript, XML, JSON, and CSS).  %
Every \m{} has an \emph{app ID}, which is a unique identifier (in a given \s{}) with random alphanumeric characters, often used for the \m{}'s requests to the \s{} server. 
Every \m{} developer has a random secret (in WeChat and Baidu, called \emph{\a{}}, 32-character long), assigned to their account, which is used to authenticate the developer of \m{} for calling the \s{} server-side APIs. The \a{} is regarded as a sensitive piece of information. 
The \emph{developer server} is the backend server of a specific \m{}, set up and maintained by the \m{} developer (not controlled by the \s{} platform). In the case of WeChat, the developer server must have a valid Internet Content Provider (ICP~\cite{icp}) licensed  domain name.
Without specifically mentioning the server or client/package, hereafter by \emph{\m{}} we refer to the entire \m{}, any code representing the corresponding developer or business.

\subhead{Access token}
To use a \s{}'s server-side APIs, and access resources associated with a \m{}, %
an \emph{access token} %
is required. This is an ephemeral secret (valid for 2 hours for WeChat and 30 days for Baidu, renewable anytime), issued by the \s{} server through an API call, e.g., \texttt{getAccessToken}~\cite{accessToken} in WeChat, which requires an app ID and \a{} as request parameters.

\subhead{OpenID}
In WeChat and Baidu \m{}s, the \emph{OpenID} is a user identifier that is unique for each \m{}. When a user logs into a \m{}, the \m{} sends a request to the \s{} server to obtain the user's openID, which is based on the user's \s{} account. In WeChat, it is an encrypted value of the user's WeChat ID and the app ID of the \m{}, and remains the same for the same user-\m{} combination. %

\subhead{WeChat cloud base functionalities}
\label{sec:cloudbase-bg}
\M{}s can take advantage of the cloud base~\cite{cloudbase}, an option
in WeChat that enables the \m{}s to utilize some basic cloud functionalities without setting up a dedicated server. 
The cloud base has a range of features, e.g., cloud functions, databases, storage and cloud call.
A \emph{cloud function} %
allows developers to execute their server-side (JavaScript) code. 
These functions can be typically triggered by specific events, such as a user action or any change in data. Within the \m{}, cloud functions can be triggered with the \m{}'s regular API (termed as JSAPI) \texttt{wx.cloud.callFunction}. 
The cloud base also offers the \emph{cloud call} capability to call the server-side APIs from cloud functions. This is recommended by WeChat apart from  
calling the server-side APIs using developer server or Tencent cloud hosting~\cite{wxcallcontainer}. Cloud calls are implicitly authenticated (no need to supply the \a{} or access token). 
The JSON \emph{cloud base database} can be queried by the \m{} to retrieve or update data using cloud functions. %
This database can be called using either JSAPIs, or the server-side APIs from the developer server (with an access token). 
The \emph{cloud base storage} provides a
storage space for \m{}s to store their files, accessible by %
dedicated APIs, cloud functions, \mbox{or the developer server.}

\subhead{WeChat IP whitelisting}
The calls to server-side APIs can be restricted to originate from only a list of IP addresses (no domain names), configured at the WeChat \m{} developer portal. If enabled by the developer (disabled by default), only these IP addresses can call the server-side APIs, i.e., no other hosts can obtain access tokens even if they have valid \a{}s.
Note that IP whitelisting applies to all the server-side APIs (i.e., not for specific APIs). 

\subhead{WeChat plug-in}
A WeChat \m{} plug-in is a package of pre-built custom components or libraries that can be integrated into a \m{}. 
Developers can request to integrate a plug-in (developed by WeChat and third-parties) from the developer portal. %

\begin{figure}
    \includegraphics[trim={0 0 0 2cm},clip, width = 0.5\textwidth]{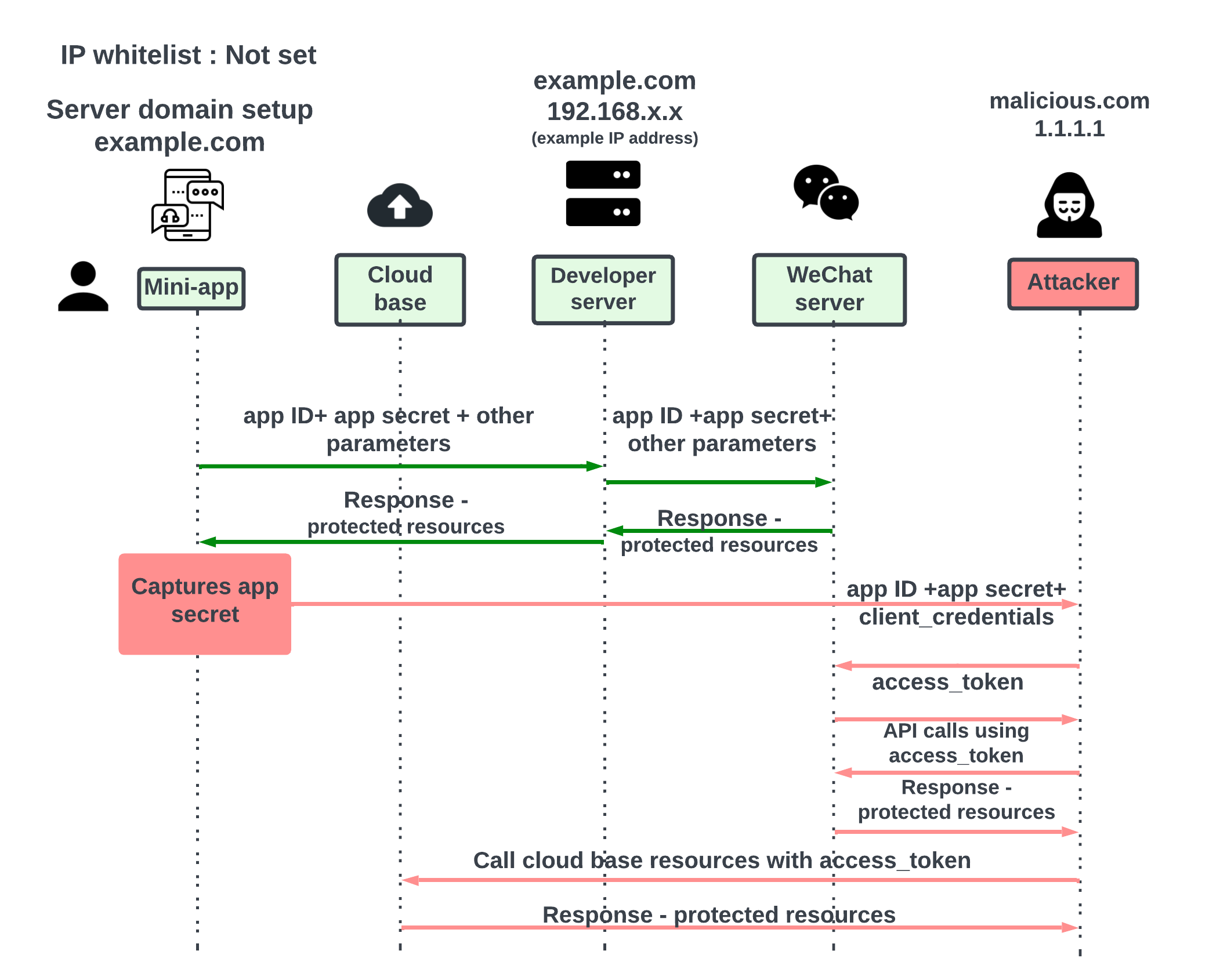}
    \caption{Example scenario of an attacker making use of the hard-coded \a{}s, by reverse-engineering the \m{} binary, and using them  to successfully access WeChat's server-side APIs.}
    \label{fig:notrecom-api}
\end{figure}

\section{Objectives and Threat Model}
\label{sec:objectives}

In this section, we formulate the secret leakage problem we consider, state our objectives, and define the threat model.

\subhead{Authentication Secret leakage problem} %
\M{}s are quite different from regular mobile apps, in terms of their reliance 
on the \s{} platform as a runtime authority for authentication, access control and other services through the server-side APIs.
There are mainly three types of authentication involved in the \s{} environment: 
\emph{\m{} user to \s{}}, when the user initially logs into the \s{} (usually persistent across reboots as in WeChat); \emph{\m{} user to \m{}}, each time (or some cases the first time) the user uses the \m{} and clicks to consent to identity sharing; and \emph{\m{} %
to \s{}}, which is often neglected as it happens behind the scene. %
Our work deals with this last case (as the other two involve the human \m{} user which is an orthogonal authentication problem). 
There exists also \emph{\m{} to \m{}} authentication~\cite{yang2022cross} via the \s{}, which can also be considered as \m{} to \s{}.

As explained in Sec.~\ref{sec:background}, a \m{} uses its \a{} (and the app ID) to obtain an access token from the \s{} server which can be used for all subsequent server-side API calls. 
This means any improper practices enabling unauthorized calls to the server-side APIs will compromise the \m{} to \s{} authentication, leading to various security issues or outright attacks. %

The \m{} authentication secret leakage problems we consider are usually reflected in the following aspects 
(1) insecure %
but common 
practices of the \m{} developers; and (2) inherent design flaws and failure to enforce their own security guidelines by \s{} platforms. 
Taking WeChat as an example, 1) the \a{} should not be included directly within the \m{} package; 2) any API involving the \a{} as its request parameter should be called by the developer server only, not from within the \m{}; 3) IP whitelisting should be configured. \M{}s developed by not following one or multiple of such recommendations can be insecure, leading to unauthorized server-side API calls; see Figure~\ref{fig:notrecom-api}. %

\subhead{Objectives}
We center our study around the issues of \m{} secret leakage and its exploitation as presented above, i.e., unauthorized calls to the server-side APIs, caused by insecure development practices.
We aim to 1) find out the extent to which 
the insecure development practices are identified from a large number of \m{}s,
despite the warnings in documentations over time;
2) analyze the feasibility of the attacker being able to actually make unauthorized calls to individual server-side APIs, for \m{}s with the identified insecure practices;
3) understand the security consequences of such unauthorized server-side API calls, in a given \s{} platform.

\subhead{Threat model}
We assume that the \m{}s are benign. 
The (source) code of the \m{} is integrity-protected by the \s{}. Also, the developer servers as well as the \s{} server used by the \m{}s are trusted and the communication between the \m{} (via the \s{}) and its developer server is through HTTPS, and is thus assumed to be secure. These assumptions are in line with the day-to-day uses of \m{}s and what has been assumed by the \m{} service providers.

\subhead{Attacker requirements and capabilities} An attacker with a regular \s{} account %
can obtain (sometimes on a large-scale) the binary package of any \m{} that is publicly visible.
This can be achieved by several means, e.g., installing the \s{} on a rooted device or its PC client with the help of some reverse-engineering tools (e.g., Frida~\cite{frida}). Then, certain open-source scripts (e.g.,~\cite{unpacker,decryptor} for WeChat) can be used to extract/unpack the content (JS and resource files). The attacker can read and change the code of the reverse engineered \m{} locally, but will not be able to 
re-publish it for the same app ID (enforced by the \s{}).  
The attacker can view the code of the \m{}, but will not have access to the \m{}'s cloud base. They can thus read the \m{}'s code for hard-coded secrets, and other information such as business logic, and use the obtained information to attack the \m{} in different ways. To access the developer server-side APIs and \s{} server-side APIs, the attacker does not need to possess any special privileges, or even a regular \s{} account (e.g., WeChat or Baidu). They just need access to an OS terminal capable of dealing with web requests or a REST client such as Postman~\cite{postman}, if the IP whitelisting for the \m{} is disabled.

\subhead{Scope}
Our study is primarily focused on WeChat \m{}s, with certain analysis extended to Baidu \m{}s. WeChat's other open platform features such as official accounts, WeChat's SSO, mobile and web development are out-of-scope. 
Other types of authentication which do not involve the explicit usage of \a{}s, e.g., \m{} user to \s{} (e.g., the attacker being able to log into someone's WeChat account), \m{} user to \m{}, and \m{} to \m{} (e.g., one \m{} impersonating another, see~\cite{yang2022cross}), are excluded.
We only examine the (reverse-engineered) \m{} packages, and the response to server-side API calls from the \s{}/developer server. We do not perform traffic analysis and do not consider JSAPIs, and other APIs, e.g., Tencent's cloud hosting APIs, third-party APIs, WeChat's payment APIs, and Baidu's cloud APIs.

\begin{figure*}[htb]
    \includegraphics[width = 0.65\textwidth]{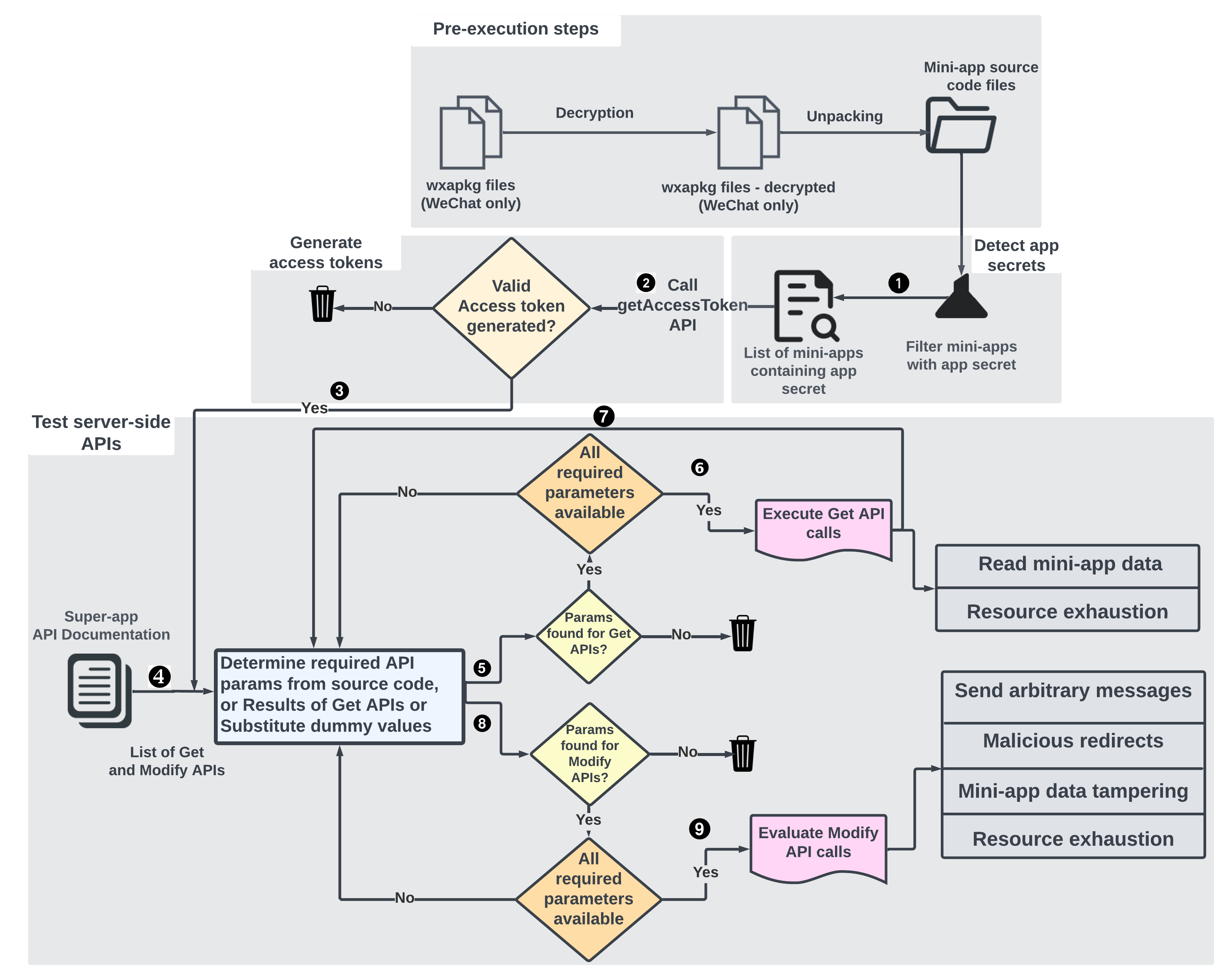}
    \caption{Overview of our analysis methodology for WeChat and Baidu \m{}s.}
    \label{fig:overview}
\end{figure*}

\section{Methodology}
\label{sec:implementation}

In this section, we discuss our methodology for conducting a step-wise analysis of WeChat \m{}s for potential unauthorized server-side API calls. The same steps apply to Baidu \m{}s as well, except for decrypting and unpacking, and IP whitelisting validation. %
Our methodology is designed to facilitate the analysis of \m{}s from other platforms that have a similar \m{} directory structure (like WeChat and Baidu) and rely directly on secret-based \m{} authentication. 
A large-scale measurement study on WeChat and Baidu using this methodology 
is presented in Sec.~\ref{sec:measurement}. We build our tool following this methodology, which is fully automated, except for separating the Get vs.\ Modify APIs from the \s{} documentations; we manually label these APIs to avoid calling the Modify APIs, which may interfere with the \m{}s' functionality. %

\subhead{Overview}
The analysis is composed of the following steps: preparation of \m{} files, detection of insecure development practices, reviewing the list of server-side APIs, and testing of potential unauthorized server-side API calls; see Figure~\ref{fig:overview}.
We consider the following insecure development practices: hard-coded \a{}s in the \m{} package,
absence of IP whitelisting, and direct invocation of server-side APIs.
First, we decrypt and unpack the WeChat \m{}s (using~\cite{decryptor} and \cite{unpacker}, respectively). This step is not needed for Baidu \m{}s. 
Then, with static code analysis of the unpacked \m{}s, we detect hard-coded \a{}s and direct invocation of server-side APIs.
Next, we use the identified \a{}s to obtain the access tokens from the \s{} server in preparation for next steps.
Meanwhile, we review (once per \s{}) the list of server-side APIs, categorize them and analyze the requirements for invoking them.
When it comes to testing potential unauthorized server-side API calls, we make use of the access tokens from the previous step to evaluate the callability of these APIs by an attacker, which at the same time also checks if IP whitelisting is enabled or not.

\subhead{Detection of hard-coded secrets}
We first manually analyzed 100 WeChat \m{}s, 
and observed that although such \a{}s are eventually used as request parameters to standard API calls (e.g., code2session API, see Appendix~\ref{wxlogin} and Figure~\ref{fig:code-snippet1}), 
they are also used to make calls to custom APIs of a \m{} developer server (see Figure~\ref{fig:code-snippet-api2} in Appendix), or even a few obfuscated functions, which will invoke the standard APIs on the server side.
To find and capture the \a{} patterns, we designed
regular expressions for WeChat and Baidu 
to identify such secrets in the unpacked \m{} code (following similar approaches based on pattern, keywords, and entropy~\cite{sinha2015detecting, meli2019bad, gitsleaksecret}). 
Some parts of the source code of the reverse-engineered \m{} are highly obfuscated, making it difficult to analyze. However, the variable names used for identifying the app ID and \a{} are not obfuscated  as observed in our initial manual analysis. Any JSON data, including API request parameters, will not be obfuscated as the server interprets the received data depending on the variable names.
Similarly, static string matching is also used against a set of 84 server-side APIs in total to detect calling of these APIs directly from the \m{}. %
Note that we do not distinguish between the case where a \m{}  discloses its own secret and the case where it discloses the secret of another \m{} (which may happen due to code cloning). All our identified \a{}s are validated by calling the \texttt{getAccessToken} API~\cite{accessToken} (see the  discussion below about obtaining the access token), ensuring that all the identified secrets are valid and there are no false positives.

\subhead{Preparing the list of server-side APIs}
We study WeChat and Baidu \m{}'s server-side API documentation~\cite{api-class,api-classv2,baiduapilist} and divide these APIs into two categories: \emph{Get APIs} that only read information, and \emph{Modify APIs} that can cause updates (Table~\ref{api-list-tested} and Table~\ref{baidu-api-list-tested} in Appendix). %
The category (Get/Modify) is determined based on the API's  intended operation and its request parameters. To further determine their callability without actually invoking them, we manually analyze the request parameters of each API and confirm if all the required request parameters can be obtained. Only when an API is confirmed to be callable, we include it in our analysis. In our measurement study, finding these request parameters and verifying the callability of Modify APIs are carried out using our automated framework. It can be inferred from Table~\ref{api-list-tested} that most of the Modify APIs can be called with the help of the response returned from the corresponding Get API, which refers to the special subset we need to call. To evaluate this predetermined list of server-side APIs in both WeChat and Baidu, we use the classification specified by the respective \s{}s. In most cases, all APIs that belong to one category are used for the same \m{} feature. Thus, we can easily assess a Modify API using the response of a Get API from that same category, and identify features susceptible to attacks.

\subhead{Obtaining the access token and the presence of IP whitelisting}
To prepare for testing the server-side APIs, as well as to 
verify the validity of the detected hard-coded secret from a given \m{}, we need to call the access token generation API. If the API returns a valid access token, it can be inferred that the \m{} is still actively present in the respective \s{} platform. We use the app ID and \a{} parameters from the static analysis along with the authorization grant type 
parameter set to a constant value \texttt{"client\_credential"}. The corresponding \s{} server verifies the app ID and \a{} and returns a valid access token. If the app ID or \a{} is incorrect, the API will return an error message which confirms the invalidity of the credentials. Note that an \a{} confirmed to be invalid presently does not mean it was invalid at the time of being hard-coded in the \m{} package.

Being able to obtain an access token in the previous step can also  
confirm the absence of IP whitelisting for a \m{} (applicable only for WeChat \m{}s). 
If the \a{} is invalid, the response from the server indicates that the app ID or \a{} is invalid; if IP whitelisting is enabled, the response mentions that the IP address of the request origin is not in the whitelist. 
We use such explicit error messages to check the adoption of whitelisting among all the WeChat \m{}s, even when we do not have the corresponding \a{} (in which case, we use a random 32-character value). %

\subhead{Testing the server-side APIs}
Using the valid access tokens obtained from the previous step, we attempt to call a subset of the Get server-side APIs that do not involve any form of modification of the \m{} data or operations (listed in Tables \ref{api-list-tested} and \ref{baidu-api-list-tested}). Calling this subset is necessary to enable analysis for the Modify APIs, and we only retrieve metadata to check if the required request parameters for calling other APIs are available without storing any data returned in the response. %
Note that this does not require possessing a WeChat or Baidu account. 
All the available server-side API requests have a calling quota and hence, we avoid making arbitrary API calls. %
If a Get 
API is successful for a particular category, then, we determine if the Modify API from the same category can be successfully called or not.
If a Get API call returns a success response, our framework can automatically determine the callability of a corresponding Modify API. This is done by making use of a combination of: the response returned by the Get API, inserting attacker-controlled dummy inputs 
(based on the API documentation), searching the \m{} source code for pertinent 
values.
We confirm the callability of a Modify API if all the required parameters are available.
Whenever we need any verification of the behaviour of Modify APIs, or resource-exhaustive calls to the Get APIs, we test them only with our own \m{}.

\section{Measurement}
\label{sec:measurement}

To achieve the three objectives put forward in Sec.~\ref{sec:objectives}, %
we quantify what can be observed from a large number of \m{}s, and for individual APIs in their respective contexts. Also, we consider the historical aspect of such insecure development practices since the extent (prevalence) we measure can also be temporal, e.g., as the warnings or recommendations in the documentation~\cite{securitydoc} can be dated back to the early days of \m{}s, what has been the prevalence over time?

\subsection{Datasets}
An early collection of WeChat \m{}s we were able to obtain is a dataset of 115,392 WeChat \m{}s (\texttt{"wxapkg"} files), crawled between July 2021 and Dec.\ 2021 (\texttt{DATASET1}).
We also used a dataset of 10,000 Baidu \m{}s crawled in Jan.\ 2023 (\texttt{DATASET2}), 
10,000 additional WeChat \m{}s, collected on Jan.\ 25, 2023 (\texttt{DATASET3}). 
These datasets were crawled using MiniCrawler~\cite{minicrawler}. 
Note that this choice of datasets can naturally cover an important case where older versions of a \m{} hard-coded the \a{} which got leaked but the \a{} still remains valid now, enabling attacks. Then, it is possible that by examining the current version of the same \m{} package, no hard-coded \a{} can be detected, hence not drawing attention 
and misleading both the developers and WeChat into a false sense of security.
For this same-\m{} temporal comparison, we also randomly picked 100 \m{}s from \texttt{DATASET1}, and analyzed them twice -- in Jan.\ 2023 and in Mar.\ 2023, before and after our report to WeChat respectively (more in Sec.~\ref{temporalcomp}). We analyzed \texttt{DATASET1}  in Dec.\ 2022 and \texttt{DATASET2} in Feb.\ 2023.

 After decrypting the WeChat \m{}s files from \texttt{DATASET1}, we ended up with 115,244 successfully decrypted \m{}s, consuming 265 GB disk space. 
From these \m{}s, we could successfully unpack 110,993 in total (434 GB in size), which we used for further analysis. 
For Baidu \m{}s in \texttt{DATASET2}, we extracted the zip files and used them as is (59 GB in size, no decryption or unpacking needed).
Among the recent 10,000 WeChat \m{}s (\texttt{DATASET3}), we managed to decrypt and unpack 9,994 \m{}s and 9,824 \m{} respectively. For an overview of our datasets and the corresponding \a{}s and generated access tokens, see Table~\ref{tab:miniappresult}.  

\begin{table}[htb]
\centering
\begin{tabular}{l|rrr}
\toprule
\textbf{\M{}s} 
& \textbf{\shortstack{DATASET1}}& \textbf{\shortstack{DATASET3}}&
\textbf{\shortstack{DATASET2}}\\
% \midrule
% \rowcolor{gray(x11gray)}
\# total &  115,392 & 10,000 &10,000\\
\# decrypted & 115,244 & 9,994& -\\
% \rowcolor{gray(x11gray)}
\# unpacked & 110,993 & 9,824 &-\\
%\shortstack{\#  of \m{}s\\with hard-coded\\\a{}s}
\# hard-coded secrets
& 43,377 & 2,894  &112 \\
% \rowcolor{gray(x11gray)}
%\shortstack{\#  of generated\\access tokens} 
\# access tokens 
& 36,425 & 2,572 &112 \\
\bottomrule
\end{tabular}
\vspace{-10pt}
    \caption{WeChat (DATASET 1 and DATASET 3) and Baidu (DATASET 2) \m{} datasets used in our measurement. ``\#~hard-coded secrets'': the number of unique \m{}s with hard-coded \a{}s; ``\#~access tokens'': \m{}s where access tokens were generated successfully.}%
    % \am{Add percentages}
  
     % \supraja{we currently don't have this type of stats for DATASET2 - baidu. I will add it here in another column or as a separate table below this?} 
    % \V{Don't add another table. Just a column. But what about whitelisting?}
    % \mm{Instead of valid secrets, we should use access tokens}}
    \vspace{-15pt}
    \label{tab:miniappresult}
\end{table}

\newcommand{\STAB}[1]{\begin{tabular}{}#1\end{tabular}}
\begin{table*}[t]
%%\small
\centering
\resizebox{\linewidth}{!}{
{\tiny
\begin{tabular}{l|llrccccl}
\textbf{Server-side APIs}&
\textbf{Required parameters} &
\textbf{\# miniapps}&
%Rotate------------------------
% \begin{rotate}{60}\textbf{[A]} \end{rotate}&
% \begin{rotate}{60}\textbf{[B]} \end{rotate} &
% \begin{rotate}{60}\textbf{[C]} \end{rotate} &
% \begin{rotate}{60}\textbf{[D]} \end{rotate} & 
% \begin{rotate}{60}\textbf{[E]} \end{rotate} & 
% \textbf{Impact}\\ 
%Rotate------------------------

%Normal------------------------
\textbf{[A]} &
\textbf{[B]} &
\textbf{[C]}  &
\textbf{[D]}  & 
\textbf{[E]} & 
\textbf{Impact}\\ 
%Normal------------------------

%WordWrap------------------------
% \textbf{\shortstack{Read Miniapp\\Data}}
% \textbf{\shortstack{Send Arbitrary\\Messages}}
% \textbf{\shortstack{Miniapp\\Tampering}}
% \textbf{\shortstack{Malicious\\Redirects}}
% \textbf{\shortstack{Resource\\Exhaustion}}
% \textbf{Impact}\\ 
%WordWrap------------------------

\hline

\textbf{clearQuotaByAppSecret}& appID, appSecret  &36,425 &  &   &\textcolor{red}\checkmark   &      &\textcolor{red}\checkmark & High \\%\hline
\textbf{clearQuota}          & AT, appID &36,425    &   &   & \textcolor{red}\checkmark  & &\textcolor{red}\checkmark  & High \\ %\hline
\textbf{managePlugin}            & AT, pluginAppID  &7,242   &\textcolor{red}\checkmark   &   & \textcolor{red}\checkmark   & &\textcolor{red}\checkmark  & High \\ %\hline
\textbf{deleteNearbyPoi}               & AT, poiID  &2,927  &   &   &\textcolor{red}\checkmark  &  &\textcolor{red}\checkmark & High \\ %\hline
\textbf{setShowStatus}           & AT, poiID  &2,927   &   &   & \textcolor{red}\checkmark    &  &\textcolor{red}\checkmark & High \\ %\hline
\textbf{managePluginApplication} & AT, appID  &772  &\textcolor{red}\checkmark   &   & \textcolor{red}\checkmark   &   &\textcolor{red}\checkmark & High\\ %\hline
\textbf{invokeCloudFunctions}   & AT, cloudFunctionName  &202  &\textcolor{red}\checkmark  & \textcolor{red}\checkmark  & \textcolor{red}\checkmark   & &\textcolor{red}\checkmark  & High \\ %\hline
\textbf{databaseCollectionGet}   & AT, cloudEnv &24  & \textcolor{red}\checkmark  &   &  & &\textcolor{red}\checkmark & High \\ %\hline
\textbf{databaseCollectionAdd}   & AT, cloudEnv, CollectionName  &24  &   &   & \textcolor{red}\checkmark  & &\textcolor{red}\checkmark & High \\ %\hline
\textbf{databaseCollectionDelete}&AT, cloudEnv, CollectionName &24  &   &   &\textcolor{red}\checkmark &&\textcolor{red}\checkmark & High \\ %\hline
\textbf{databaseAdd}             & AT, cloudEnv  &24  &   &   &\textcolor{red}\checkmark  & &\textcolor{red}\checkmark & High \\ %\hline
\textbf{databaseDelete}          & AT, cloudEnv  &24  &   &   & \textcolor{red}\checkmark   & &\textcolor{red}\checkmark & High \\ 
% \hline
\textbf{databaseUpdate}          & AT, cloudEnv  &24  &   &   & \textcolor{red}\checkmark & &\textcolor{red}\checkmark & High \\ 
% \hline
\textbf{databaseQuery}           & AT, cloudEnv  &24   & \textcolor{red}\checkmark  &   &\textcolor{red}\checkmark    & &\textcolor{red}\checkmark & High \\ 
% \hline
\textbf{setUpdatableMsg}         & AT  &17   &   & \textcolor{red}\checkmark   & \textcolor{red}\checkmark   & \textcolor{red}\checkmark&\textcolor{red}\checkmark   & High\\ 
% \hline

\textbf{uploadTempMedia}         & AT  &36,425  &    &  & \textcolor{red}\checkmark   &  &\textcolor{red}\checkmark &Medium \\ 
% \hline
\textbf{getApiQuota}          & AT, cgi\_path  &36,425  & \textcolor{red}\checkmark  &   &   & &\textcolor{red}\checkmark  & Medium \\ 
% \hline
\textbf{getDomainInfo}          & AT  &33,795  & \textcolor{red}\checkmark  &    &   & &\textcolor{red}\checkmark  & Medium \\ 
% \hline
\textbf{getFeedback}          & AT &18,475 &\textcolor{red}\checkmark  &   &   & &\textcolor{red}\checkmark  & Medium \\ 
% \hline
\textbf{customerServiceMessage.send}  & AT, openID   & 18,224&   & \textcolor{red}\checkmark  &  & \textcolor{red}\checkmark &\textcolor{red}\checkmark & Medium\\ 
% \hline
\textbf{getQcloudToken}          & AT &11,786   &\textcolor{red}\checkmark &     &   & &\textcolor{red}\checkmark  & Medium \\ 
% \hline
\textbf{getAllDelivery}              & AT  &8,622   & \textcolor{red}\checkmark  &     &   & &\textcolor{red}\checkmark & Medium \\ 
\textbf{getPrinter}              & AT  &312   & \textcolor{red}\checkmark  &     &   & &\textcolor{red}\checkmark & Medium \\ 
% \hline
\textbf{updatePrinter}           & AT, openID  &312   &   &   & \textcolor{red}\checkmark   & &\textcolor{red}\checkmark  & Medium \\ 
% \hline

\textbf{createActivityId}        & AT  &36,425   &\textcolor{red}\checkmark   &  &   &  &\textcolor{red}\checkmark & Low \\ 
% \hline
\textbf{getNearbyPoiList}         & AT  &2,927   & \textcolor{red}\checkmark  &     &   & &\textcolor{red}\checkmark  & Low \\ 
\hline

\textbf{addTemplate}*          & AT  &112  &   &   & \textcolor{red}\checkmark  & &\textcolor{red}\checkmark & Medium \\ 
% \hline
\textbf{submitResource}*          & AT &112  &   &   &\textcolor{red}\checkmark   & &\textcolor{red}\checkmark  & Medium \\ 
% \hline
\textbf{submitSitemap}*          & AT &112   &  &   & \textcolor{red}\checkmark  & & \textcolor{red}\checkmark & Medium \\ 
% \hline
\textbf{interfaceSubmission}*          & AT   &112  &   &   & \textcolor{red}\checkmark   & & \textcolor{red}\checkmark & Medium \\ 
% \hline
\textbf{submitsku}*          & AT  &112 &  &   &\textcolor{red}\checkmark & &\textcolor{red}\checkmark  & Medium \\ 
% \hline
\textbf{createCoupon}*          & AT  &112  &  &   & \textcolor{red}\checkmark    & & \textcolor{red}\checkmark & Medium \\ 
% \hline
\textbf{submitcoupon}*          & AT  &112  &  &   &\textcolor{red}\checkmark    & & \textcolor{red}\checkmark & Medium \\ 
% \hline
\textbf{ManageCoupon}*          & AT  &112   &  &   & \textcolor{red}\checkmark    & & \textcolor{red}\checkmark & Medium \\ 
% \hline
\textbf{getTemplateList}*          & AT &74  & \textcolor{red}\checkmark  &   &    & &\textcolor{red}\checkmark  & Medium \\ 
% \hline
\textbf{deleteMessageTemplate}*          & AT  &74 &   &   & \textcolor{red}\checkmark  & &\textcolor{red}\checkmark  & Medium \\ \hline
        \end{tabular}}}
\vspace{-10pt}
\caption{Statistics of unauthorized callable WeChat (\texttt{DATASET1}) and Baidu (\texttt{DATASET2}) server-side APIs, including their required parameters. 
% Since \texttt{invokeCloudFunctions} calls cloud functions are implemented by developers, the actual impact vary a lot depending on what the function does \am{Why this is needed in the caption?}. 
Items marked with * denote Baidu APIs (at the lower part of the table). 
[A]: Read \M{} Data; [B]: Send Arbitrary Messages; [C]: Data Tampering; [D]: Malicious Redirects; [E]: Resource Exhaustion; AT: Access Token; \textcolor{red}{\checkmark\ } denotes the possibility of the attack using the corresponding server-side API. Impact: the impact of the attacker\textquotesingle s invocation of the API on a \m{} and its users --- determined based on the CVSS calculator %~\cite{cvsscalc}
(see Appendix~\ref{cvssdetails}).
% \mm{add something about Invocation and Impact; also change color for xmarks}
}%
%\vspace{-15pt}
\label{tab:api-imapct}
%\vspace{-10pt}
\end{table*}

\subsection{Measurement Results}
\label{sec:results}

\subsubsection{Insecure Development Practices} 
\subhead{\\Hard-coded \a{} and IP whitelisting} 
Our regular expressions to match the \a{}s resulted in hits in 43,337 out of the 110,993 (successfully decrypted and unpacked from \texttt{DATASET1}) WeChat \m{}s (about 39\%), and 112 out of the 10,000 Baidu \m{}s (about 1\%). 
When we attempted to generate access tokens from these \a{}s, we found a total of 36,425 (approximately 33\%) WeChat \m{}s with valid hard-coded secrets (i.e., successfully generated access tokens), violating the security guidelines by WeChat in terms of both hard-coding the \a{}s and not configuring the IP whitelisting. On the other hand, all the Baidu \m{}s with hard-coded \a{}s generated valid access tokens, meaning all the identified secrets are valid. %

The access token API returned failure responses for 6,959 WeChat \m{}s, in which for 3,578 \m{}s, the \a{} was invalid. For another 3,374 \m{}s, the API returned a \texttt{50002} error code stating that ``the user is limited.'' 
We did not find enough information about this error code, except that, 
as per the documentation, the user is not authorized to use this API~\cite{error-code}. For the remaining 7 \m{}s, the API returned that the requesting IP address is not in the list of whitelisted IPs. 
The use of IP whitelisting is indeed very limited as we found out by testing all the 110,993 WeChat \m{}s, using dummy \a{}s for the ones without hard-coded secrets, that only 33 \m{}s have IP whitelisting configured. 

\subhead{Direct invocation of server-side APIs}  \label{directinvocation}
Attempting to call server-side APIs within the \m{} directly is not recommended by WeChat or Baidu, not because the direct invocation itself causes security issues, but because to make such calls successful the developer must involve both hard-coding \a{}, and disabling IP whitelisting (in WeChat), as a \m{} client can be run from any IP address. Therefore, we checked the prevalence of such practice.
We found 4,098 occurrences of direct invocations of server-side APIs from 2,317 \m{}s out of the 110,993 unpacked WeChat \m{}s. We further classify each API call based on the category~\cite{api-class,api-classv2} as shown in Table~\ref{tab:apidirect} in Appendix. We detected a much lower number of direct invocations in Baidu \m{}s compared to WeChat. In total, there are 43 occurrences of direct server-side APIs invocations in 11 \m{}s in Baidu (mostly \texttt{getSessionKey} and \texttt{getTemplateList} APIs). %

\subsubsection{Unauthorized Invocation of Server-side APIs} \label{sec:callabilityw}
With a valid access token, all the server-side APIs should be callable for a given \m{}. However, in practice, each API has its semantics and the \m{}'s functionality and current state determines whether a specific API can be called or supported at a given time. Therefore, next, we examine individual APIs' callability for the chosen server-side APIs under each category (see Tables~\ref{api-list-tested} and ~\ref{baidu-api-list-tested},
the category names are from the official WeChat and Baidu documentation~\cite{api-class, baiduapilist}).
We present the statistics of the successful server-side API calls, based on the chosen 26 WeChat server-side APIs, 
and 10 Baidu server-side APIs we tested in Tables~\ref{tab:api-imapct}, and Table~\ref{tab:api-imapct-new} (in appendix). We also discuss selected per-category results for WeChat below (see Appendix~\ref{sec:callabilityb} for Baidu).

\begin{sloppypar}
\subhead{Customer service messages} We test this category of APIs to determine if an attacker can send messages to \m{} users and insert arbitrary media to user messages. For this category, 
we evaluate two APIs, \texttt{uploadTempMedia}, and \texttt{customerServiceMessage.s-\\end}. We observe that if the Customer Service Message feature is enabled for a \m{}, then \texttt{uploadTempMedia} is callable, as it requires only the media file, form data and the access token as its request parameters. 
The \texttt{customerServiceMessage.send} API is callable %
only when the openID is present. Our framework identified that openIDs are disclosed for 312 (<1\%) \m{}s via the \texttt{getPrinter} API, and for 18,475 (50.7\%) \m{}s via the \texttt{getFeedback} API; hence  customer service messages can be forged for such \m{}s. 
Here, we do not consider other potential sources for obtaining openIDs such as cloud functions which may depend on the \m{}'s business logic. \looseness=-1
\end{sloppypar}

\subhead{Cloud base HTTP API}  
If the API \texttt{invokeCloudFunction} is callable,
for the \m{}s containing the access token and the cloud function calls in the \m{} code, our framework statically searched through each \m{} directory to identify the cloud function calls and collect their names. Among the total 36,425 \m{}s that we tested, 254 (<1\%) \m{}s use cloud base, and our framework collected 202 (<1\%) \m{}s with cloud functions having valid access tokens, containing a total of 992 distinct cloud functions. 
As several cloud functions involved update, delete operations in the cloud level, we did not invoke any of the collected cloud functions. %

 We also test the cloud database CRUD (Create, Read, Update, Delete) APIs to verify if an unauthorized access to the database is possible. Out of the total 11 database APIs available, we directly test only one of them -- \texttt{databaseCollectionGet}, which returns only the table names from the cloud database. With these names, we
only evaluate the remaining 6 chosen APIs related to cloud database based on their callability. Through statically searching the \m{} source code, we see that out of 254 (<1\%) \m{}s that use cloud base, 179 (<1\%) \m{}s use the cloud database. We identify only for 24 (<1\%) \m{}s out of 179 \m{}s database APIs are callable, as the API requires cloud environment ID as its request parameter. The remaining 230 (<1\%) \m{}s either did not have a valid cloud environment ID in the \m{}'s code or the APIs returned with an error stating that the the number of requests exceeded the quota for the \m{}. We further evaluated other database APIs considering the 24 \m{}s. We find that the \texttt{databaseAddCollection} and \texttt{databaseDeleteCollection} can be performed for all the 24 \m{}s. Additionally, adding a record to the database collection, updating and deleting are also possible for these 24 \m{}s.  We do not attempt to access or download any of the identified database tables apart from evaluating the callability by assessing the request parameters for the corresponding server-side APIs.
 
 We then test for the Tencent cloud credentials API \cite{tencentcloudapi}. Out of the total number of \m{}s tested, we obtain the Tencent cloud API calling credentials for 11,786 (32.3\%)  \m{}s. For the remaining 24,639 (67.6\%) \m{}s, the QCloud token API returned an error message stating that the \m{} has no cloud base privilege, meaning that the \m{} does not use Tencent cloud base. We did not test the cloud storage APIs, as when evaluated, we see that the upload link, download link and batch delete APIs require the file path and value where, in most cases, it is unknown to an unauthorized user. If the storage paths are revealed in some part of the code, it is easier to upload arbitrary files, download files and delete files from the cloud storage. In our analysis, we did not find any hard-coded storage path in the tested \m{}s.

 \subhead{Plug-in management} We test this category of APIs to verify if an attacker will be able to misuse the plug-in-related functionalities of a \m{}.  The API \texttt{getPluginList} (\texttt{managePlugin} with \texttt{"list"} as parameter in v2 API documentation) is to retrieve the current in-use plug-ins for any \m{}. No error code was received for 17,433 (47\%) \m{}s when calling \texttt{getPluginList} API. For the remaining 18,992 (52.1\%) \m{}s, the API returned with an error 
stating that accessing the API is unauthorized, meaning that the \m{} has not configured plug-in related permissions in the developer portal. Since the plug-in app ID is public information for any plug-in, the \texttt{applyPlugin} (\texttt{managePlugin} with "apply") API is callable for these 17,433 \m{}s, as identified by our framework. This API takes in only the access token and the plug-in app ID as request parameters. For 7,242 (19\%) \m{}s that make use of the plug-ins (returning a non-empty list), the \texttt{unbindPlugin} (\texttt{managePlugin} with "unbind") API will be callable. %

\subhead{openAPI management} We verify openAPI management APIs to determine if the API management details can be obtained and modified by the attacker. 
We selected one Get API \texttt{getAPIQuota} and two Modify APIs \texttt{clearQuota} and \texttt{clearQuotaByAppSecret}. Using our analysis framework, we make calls to the \texttt{getAPIQuota} API for all the \m{}s with valid access tokens. The \texttt{clearQuota} API and \texttt{clearQuotaByAppSecret} API can also be called for the \m{} with valid hard-coded \a{}s and access tokens generated.
Therefore, the attacker can view the used quota for every server-side API and reset the quota at any time.

\subhead{\M{}s nearby} This is a feature for business \m{}s to show up in WeChat (under \m{}s nearby) when a user is in proximity to their business location. We test this category to check if an attacker will be able to add or modify the nearby points of interest of a \m{}. 
We found 4,918 \m{}s (13\%) had this feature enabled (via \texttt{nearbyPoi.getList}); 2,927 (8\%) \m{}s had configured some POIs, for which an attacker will be able to call the \texttt{deletePOI} and \texttt{setShowStatus} APIs to delete or change the visibility of the POIs, respectively.  8,110 (22\%) \m{}s returned an error stating that the \m{}s are personal \m{}s (as opposed to business \m{}s). For 23,397 (64\%) \m{}s, the API returned an error stating that the nearby features are blocked.

\subhead{Logistics assistant} 
This category of APIs allows business \m{}s to manage logistics such as the delivery of products.
We chose 2 Get APIs (\texttt{getAllDelivery} and \texttt{getPrinter}) to test if an attacker will have access to the logistics information of a \m{}. For  26,190 (71.9\%) \m{}s, calls to \texttt{getAllDelivery} API were successful but only 8,622 (23\%) \m{}s returned the \m{} logistics delivery details as a result of this API. We conducted further tests using the \texttt{getPrinter} API, which we rely on for the openIDs to test customer message APIs. 312 (<1\%) \m{}s returned data for the \texttt{getPrinter} API, which in turn is useful in calling the \texttt{updatePrinter} API, and contains the sensitive user openIDs (required for other attacks e.g., malicious redirects). For the remaining 10,235 (28\%) \m{}s, both APIs returned an error stating that the API is unauthorized (i.e., the logistics assistant is not configured, or the \m{} is for personal use as opposed to business).

\subhead{Updatable messages} %
We test this category of APIs to see if an attacker can update the messages that are already posted to users. We call the \texttt{createActivityId} API to generate a unique activity ID which can be further used to call the \texttt{setUpdatableMessage} API. 
Although the \texttt{setUpdatableMessage} API is always callable, we find that only 17 \m{}s use this feature. %

\subhead{Operations and management} We pick two Get APIs from this category and test by calling them to verify if an attacker is able to access the \m{}'s server domain configuration and \m{}'s feedback from the users.
18,475 (50\%) \m{}s returned customer feedback as a result of the \texttt{getFeedback} API, which also returns the user's openID against each feedback. %
The openID returned by this API can further be exploited in the customer service message API to send phishing messages and malicious redirects.

\subsubsection{Testing dangerous server-side APIs with our own \m{}} 
To confirm the callability of all the server-side APIs, including the Modify APIs (which we could not call in live \m{}s of other developers), we developed our own simple WeChat \m{} and published it on the WeChat \m{} platform. With this published \m{}, we enabled all the available features provided for individual developers, and we make all the Get and Modify API calls that apply to personal \m{}s. We could call all the Get APIs successfully after generating the access token. We employed the response returned by the Get APIs to call the Modify APIs. We enabled the plug-in feature for the \m{}, added a few relevant plug-ins from the developer portal and used them in our \m{}, which we later deleted using the server-side API, breaking the entire \m{} functionality. We further accessed all the cloud database related information, and updated/deleted all our stored (test) information. As a last step, we also made continuous arbitrary calls to the server-side APIs, exhausting the API calling limit to the point that the \m{} cannot make any further calls to the exhausted server-side APIs. Our proof of concept attacks confirm the exploitability of \m{}s with known \a{}s. Since Baidu does not allow individual developer \m{}s (only enterprises), we were unable to confirm the callability of Modify APIs in
Baidu. %

\subsection{Temporal Comparison} \label{temporalcomp}
We perform two additional measurements to address another aspect of our first objective regarding the extent of the insecure development practices over time.

\subhead{Prevalence over time}
To learn the trend of the prevalence of the insecure development practices, we also conducted the same measurement on the more recent \texttt{DATASET3} in February 2023, and we found the \a{}s in 2,894 (28.9\%) \m{}s, out of which 2,572 (25.7\%) \m{}s have valid secrets in them. Our additional analysis did not change the outcomes of the evaluation that we performed with the older dataset; see Table~\ref{tab:miniappresult} for the comparison between \texttt{DATASET1} and \texttt{DATASET3}, and Table~\ref{tab:api-imapct-new} in Appendix for breakdown on individual server-side API calls and their impacts.

\subhead{Same-\m{} comparison}
We randomly selected 100 WeChat \m{}s with hard-coded \a{}s detected from the previous analysis (\texttt{DATASET1}, crawled in 2021), and downloaded their current version (on January 24, 2023 and again on March 6, 2023)
through an updated reverse-engineering approach with Frida~\cite{frida} %
(due to the technical evolution of WeChat, hooking a different library and function), and rerun our analysis for comparison.

This experiment is motivated by the fact that hard-coding \a{}s (as well as not configuring IP whitelisting in WeChat) has been discouraged or even prohibited (but not enforced), only recently more strict checks were introduced~\cite{devtoolupdatelogs}. 
So, we would like to see if there was any change to the \m{}s vulnerable to those unauthorized server-side API calls. WeChat has recently enforced 
the restriction on launching a \m{} when it has hard-coded \a{} in it (exact timeline unknown, but this was observed coincidentally after our report was filed). We verified this by attempting to submit our own \m{} with the hard-coded \a{} in plain-text. Our \m{} got rejected during the audit phase stating the obvious reason of having \a{} in the code. However, with our temporal comparison, it is evident that the \m{}s that are already on the platform with the \a{}s are still not patched, thus remain vulnerable to attacks.

Out of the total 100 \m{}s 
investigated that were identified with valid \a{}s in our initial analysis, 83 still have the same \a{}s hard-coded into their source code in the latest versions. Four have eliminated the \a{} from the source code, yet the \a{}s that were previously identified in these \m{}s are still valid (access token generated). Three \m{}s have changed their \a{}s to other values, which are also hard-coded in the current versions. Among the remaining, 9 \m{}s which had valid secrets before are currently invalid  and their current source code does not contain any \a{}s. One \m{} with hard-coded \a{} has IP whitelisting configured (previously absent). 

These results for the 100 \m{}s remained the same both in our tests in January 2023 and in March 2023, showing no very significant changes in \m{}s with hard-coded \a{}s that were published before the current (no hard-coded \a{}) enforcement by WeChat. In the end, 90/100 \m{}s remain vulnerable. 

We further tested all the previously identified WeChat \m{}s with \a{}s (36,425 \m{}s) if those \a{}s are still valid by trying to generate the access tokens again in March 2023. We find that out of the total tested, for 36,293 \m{}s (99.6\%) the secrets remain valid. For the ones that are no longer valid, the \texttt{getAccessToken} API threw an error stating that the app secret is invalid, except for 2 \m{} which has IP whitelisting enabled and configured. However, in our recent analysis of the same dataset in July 2023, we observed a significant decrease in the percentage of valid secrets to 16.5\% (18,332 \m{}s) as a result of WeChat's restrictions on the use of \a{}s. Out of these, 3,414 \m{}s leak the \a{}s of other \m{}s (i.e., own \a{} leakage by 14,918 \m{}s).

\subsection{Implementation and Efficiency}
We use Python to implement the automated analysis with 4,643 lines of code. The framework consists of three main components in common for WeChat and Baidu: (1) static searching of hard-coded \a{}s and validating them, (2) calling the server-side Get APIs, and (3) evaluating the Modify APIs. We also implement the cloud base APIs testing for WeChat \m{}s. Based on our observations, we created two separate regular expression patterns for WeChat and Baidu \m{}s for finding \a{}s. 
We run our analysis on an Ubuntu desktop (Intel i7-10700, 2.90GHz, 16GB RAM). For WeChat \m{}s, we first decrypted and unpacked them, which on average took 3.98 seconds/\m{}.  The average lines of JavaScript code per \m{} (both WeChat and Baidu) are approx.\ 10,000 and the average time taken for finding an \a{} is 3.58 seconds/\m{}, and the actual evaluation of one \m{} is approx.\ 6 minutes against 26 WeChat server-side APIs and approx.\ 4 minutes against 10 Baidu server-side APIs. 
To fully utilize our CPU, the analysis was carried through 7 
parallel threads, and it took about 3 days to analyze all the WeChat \m{}s, and about 5 hours for Baidu \m{}s.

In terms of effectiveness of our \a{} search via regular expressions, successfully generating the access tokens ensures the validity of the identified \a{}s, avoiding false positives. Although as we identified in 9,244 WeChat \m{}s and in 21 Baidu \m{}s with multiple hard-coded \a{}s, only one 
of those secrets turns out to be valid after the token generation attempt. To identify false negatives, we randomly chose 200 WeChat \m{}s and 50 Baidu \m{}s where no hard-coded \a{}s were identified by our regular expression, and checked them manually. Out of the 250 \m{}s, no hard-coded \a{} was found. 

\section{Security Consequences} \label{sec:consequences}
\label{sec:cases}

Based on the analysis results presented in Sec.~\ref{sec:results}, in this section, we address our third objective to understand the security consequences of the unauthorized server-side API calls caused by hard-coded \a{}s and absence of IP whitelisting.

\subsection{Consequences from Server-side APIs} 
After analyzing individual server-side APIs for their callability by an attacker based on per-\m{} semantics/states, we come up with six impact categories by reviewing several sources (e.g.,~\cite{owasp,cwehardcoded}) adapted to the \m{} paradigm; see Table~\ref{tab:api-imapct} for a complete list of the APIs tested and matched with the impact categories, and their corresponding CVSS severity scores, as we determined using the CVSS calculator (Appendix~\ref{cvssdetails}).
We provide further details on these categories below; see Table~\ref{tab:consequences} in Appendix for the statistics of \m{}s against each consequence.

\subhead{Authentication bypass and entity impersonation} 
Pertaining to the type of \m{} to \s{} authentication, we have studied the problem of unauthorized calls to server-side APIs, and how the \s{} server authenticates individual \m{}s.
This is very different from the other types of authentication where a human user is involved to provide knowledge/possession/biometrics at the time of authentication. For \m{}s to get authenticated with the \s{} server, which in some sense is on behalf of the \m{} developer, the developer needs to store a form of secret securely in advance to be used at runtime. 
The WeChat way is to recommend the use of a developer server to call their server-side APIs where the \a{} can be stored, assuming the developer server is secure (which is relatively true). However, in practice, some developers make direct server-side API calls within the \m{}, which is not blocked by WeChat or Baidu (blocked in the WeChat developer portal, but can still be unchecked in the devtool for development and testing), and hence creating the need for hard-coded secrets in the \m{} package.
IP whitelisting in WeChat was expected to add further restrictions on which server IPs can be used to make calls involving \a{}s; however, this again required developers' understanding and engagement, which does not happen readily in practice. 
In consequence, an attacker having extracted the \a{} from a package will be able to generate a valid access token from any system and effectively bypass \m{} to \s{} authentication impersonating the \m{} developer. 
\S{} servers assume that any request that contains a valid \a{} or access token is a legitimate request and respond with the requested data, and thus allow the attacker to interact with the \s{} servers. %
What makes it worse is that whoever calling these APIs do not even need to possess a \s{} account, due to no binding to the \s{} (e.g., WeChat, Baidu) environment.

\subhead{Reading \m{} data} 36,425 WeChat \m{}s from \texttt{DATASET1} and 112 Baidu \m{}s from \texttt{DATASET2} 
are subject to data exposure, through unauthorized server-side API calls (when the corresponding feature is enabled by the developer); see Table~\ref{tab:api-imapct}. 
Examples of sensitive \m{} data that can be exposed include but are not limited to \m{} data analytics, %
email addresses, security questions and answers, order IDs, tracking numbers and transaction information. This undermines data confidentiality of the \s{} ecosystem, both for the millions of users of affected \m{}s and their developers. We refrain from measuring the extent of such leakage due to obvious ethical issues.

\subhead{\M{} data tampering} 
Through unauthorized server-side API calls, in addition to losing confidentiality, \m{} data can also be tampered with. For example, when a WeChat \m{} has the plug-in feature enabled and the \m{}'s category matches with the plug-in's category, the \texttt{managePlugin} API can send arbitrary requests to the legitimate plug-in apps, and the same API can remove any plug-in from the live \m{}, thus breaking the \m{}'s functionality (details in Sec.~\ref{sec:callabilityw}). 
Unauthorized access to a \m{}'s cloud database can lead to more catastrophic consequences as an attacker can read/modify the entire database, including deleting it. This combined with the plug-in deletion can bring down the \m{} completely, thus affecting the availability of the \m{}.
Unauthorized calls to the previously mentioned \texttt{deletePOI} and \texttt{setShowStatus} can obviously manipulate configured points of interest of a business \m{}. 

\begin{sloppypar}
In Baidu, with the APIs such as \texttt{submitResource}, and \texttt{submitSitemap}, an attacker can upload arbitrary file resources to the \m{}. This unauthorized access also enables the attacker to create and submit fake coupons for the corresponding vulnerable \m{}s, thus adding arbitrary and fake data to the server.
\end{sloppypar}

\subhead{Resource exhaustion attacks}
As all the WeChat and Baidu \m{} server-side APIs have a quota limit (how many times an API can be called), the attacker can call these APIs repetitively until the limit is exhausted and are no longer available for the legitimate in-\m{} usage. Therefore, we can consider all the \m{}s for which an access token can be successfully generated to be vulnerable to resource exhaustion attacks. What is worse, in WeChat, by using the \texttt{getApiQuota} API, an attacker can obtain the quota for each API further facilitating the attacks, e.g., by making just enough number of calls to exhaust the limit; and the \texttt{clearQuota} API may also be exhausted for the legitimate user which can only be called 10 times per month. All these harm the availability of \m{}s in the \s{} ecosystem.

\iffalse
 \subhead{Arbitrary Uploads} It is seen that the uploadMedia API can be called for all the \m{}s and hence, all 36,425 \m{}s are vulnerable to arbitrary image uploads. This allows the attacker to upload malicious files with .png, .jpeg extensions that can be later fetched from the WeChat's server. However, it is to be noted that the exploitability of this attack vector solely depends upon how the image file is used in the WeChat server.
\begin{figure}[H]
    \includegraphics[width = 0.5\textwidth]{figs/Impact.png}
    \caption{Security consequences of unauthorized calling of server-side APIs}
    \label{fig:methodology1}
\end{figure}
\fi

\subhead{Sending arbitrary messages and malicious redirects} %
From Customer Service Messages in Sec.~\ref{sec:callabilityw},
it is clear that an attacker can send arbitrary customer service messages to the affected WeChat \m{} users. This message can be a text, an image, an external link or a link to open another \m{}. Attackers can obtain the openIDs of an affected \m{} with the help of other server-side APIs like \texttt{getFeedback} and \texttt{getPrinter} (also cloud functions or database, in some cases), and use those openIDs to send arbitrary messages. Further,
attackers can also modify the news-feed of the affected WeChat \m{}s as discussed in Sec.~\ref{sec:callabilityw} (under ``Updatable Messages''), including past messages from a user's feed. This can lead to fraud and phishing attacks, and malicious redirects. In Baidu, unlike WeChat, we did not find the leakage of openIDs, and hence, sending malicious messages to users seems infeasible.

\vspace{-10pt}
\subsection{Consequences from Cloud Functions} \label{cloudconsequences}
As WeChat cloud functions provide the developers with an online space to host business logic, they will not be part of the \m{} package. In the package, we only see where a cloud function call is made, the cloud function names, and the parameters passed to that function. These cloud functions can also be triggered by the developer's server using the server-side API \texttt{invokeCloudFunction}. As the API requires only the cloud function name, the parameters to that function and the \m{}'s access token, an attacker can easily use the server-side API to invoke the cloud function by using the extracted \a{}. 
Making such unauthorized calls can further harm business logic in the cloud, impacting all the \m{}'s users.

\begin{sloppypar}
As an example, we observed that an e-commerce \m{} with hard-coded \a{} has multiple cloud functions such as \texttt{confirmCustomerOrder}, \texttt{editCustomerOrder}, \texttt{changeDeliveryAddress}, \texttt{changePaymentStatus}, and \texttt{cancelOrders}, with order ID and openID as the parameters to these functions. It appears to be possible to obtain the order ID and openID by calling the \texttt{getCurrentOrderList} cloud function, which will return a list of all orders, containing the order ID, openID, order date, total cost, and delivery address. As a result, an attacker can simply use the \texttt{curl} command on the terminal to call \texttt{invokeCloudFunction} with the identified cloud function name \texttt{getCurrentOrderList}, and obtain all the currently placed orders'  order IDs and openIDs. By employing this information, the attacker can invoke other cloud functions to modify the order address to their own, cancel orders arbitrarily, edit current orders, change the payment status, etc. 
\end{sloppypar}

Another important aspect of the cloud base is that it can be shared between other services of a WeChat \m{} developer, such as other \m{}s, and WeChat official accounts that can provide a hub for branding products and gather followers. If one of their \m{}s has its secret hard-coded in the \m{} code, using that secret, an attacker can access the common cloud base (e.g., cloud functions, database, storage) and attack or even worse, take down all the relying services of that cloud base.

\section{Discussion}
\label{sec:discussion}
In here, we discuss the takeaways and reflections on our measurement and analysis of the \m{} to \s{} authentication problem due to developer secret leakage, as well as our limitations.

\subhead{Root cause analysis} %
After examining the code of the large number of \m{}s and corresponding documentations, we try to further understand why such \a{} leakage problems happen.

\emph{Binding through \s{}}.
When making calls to \s{} server-side APIs, the subject being authenticated is just \emph{code} (at least in WeChat, TikTok and Baidu) on behalf of the \m{} developer, or in certain terminologies, the third-party user or the merchant. 
Without another aiding entity, the \a{}s must be stored with the code. 
This is in contrast to the \texttt{getAuthCode} API of Alipay \m{}s~\cite{AlipayAPIDoc}, which prompts the user through the \s{} to consent and returns an authorization code (equivalent to a dynamic \a{}). This code can then be used with \texttt{applyToken} to get an access token to call server-side APIs. Access tokens generated this way will be per user session, ``endorsed''  by the \s{} (as the \s{} server would not assign a dynamic \a{} to random requesting code, unlike the static \a{} which can be pre-assigned).\footnote{The same authorization also exists in WeChat/Baidu but not for this purpose. Only when the API involves accessing \s{}-stored user information, a dynamic app secret is used along with the regular \a{} (using \texttt{code2Session} API for WeChat and \texttt{getsessionkey}~\cite{baidulogin} for Baidu), which still does not avoid hard-coding \a{}s.}
Therefore, we can see %
that involving the \s{} to locally authenticate the \m{} code and generate dynamic \a{}s can avoid having to hard-code static \a{}s, which is not the case for \s{}s like WeChat and Baidu.

\emph{Intended use of server-side APIs.}
Despite having to hard-code \a{}s as discussed above, the location matters. Hard-coding \a{}s in the \m{} is considered an insecure practice according to WeChat, and the intended way is to manage the entire business logic on the developer server where the \a{} is stored and used to make all the \s{} server-side API calls. However, our measurement clearly showed that shifting this to \m{}s (potentially falling in the wrong hands) can lead to significant security consequences.
Therefore, our observation is that \m{} developers involved insecure development practices and the \s{} platform also failed to prevent them in the first place.

\subhead{Comparing WeChat and Baidu with other \s{} platforms}
Other miniapp platforms like QQ,\footnote{QQ \m{}s - \url{https://q.qq.com/wiki/develop/miniprogram/frame/}} Duoyin (Chinese TikTok), Toutiao,\footnote{Duoyin and Toutiao from ByteDance - \url{https://developer.open-douyin.com/docs/resource/zh-CN/mini-app/introduction/overview/}} and DingTalk\footnote{\url{https://open.dingtalk.com/document/orgapp/introduction-to-dingtalk-mini-programs}} also offer \m{} features with their own server-side APIs, and similar to WeChat, they require the usage of an access token for these APIs' invocation. QQ, being under Tencent's ownership, shares substantial similarities with WeChat in terms of API functionalities~\cite{qq-mini}. Duoyin and Toutiao, developed by ByteDance, follow a similar pattern, requiring access tokens obtained through the \texttt{getAccessToken} API with parameters like app ID, \a{}, and grant type set to \texttt{"client\_credentials"}~\cite{duoyinAT}. DingTalk, developed by Alibaba, provides different \m{} platforms but retains a common access token acquisition process through the \texttt{gettoken} API, utilizing parameters like corpID and corpsecret~\cite{dingtalk-mini}. Our approach for WeChat and Baidu may also apply to the above three platforms, and lead to similar attacks if the front-end code of \m{}s from these platforms discloses app secrets, despite the documentations stating otherwise~\cite{dingtalk-mini, qq-mini, duoyinAT}. %

In contrast, Paytm,\footnote{Paytm - \url{https://business.paytm.com/docs/miniapps/overview}} a prominent digital payments app in India, utilizes a different approach for access token retrieval. The \texttt{getAccessToken} API requires 
the client ID and client secret in the request header, along with other parameters in the request body, such as the %
scope, grant type, and authorization string obtained through the \texttt{paytmFetchAuthCode} JSAPI~\cite{paytm-mini}. Similarly, Alipay\footnote{Alipay - \url{https://miniprogram.alipay.com/docs/}} \m{}s have their own set of server-side APIs, where the access token is obtained using the \texttt{applyToken} API with specific parameters like the \m{}'s ID, authClientID, grant type set to \texttt{"AUTHORIZATION\_CODE"}, and the corresponding authCode acquired through the \texttt{my.getAuthCode} JSAPI~\cite{AlipayAPIDoc}. It is worth noting that the acquisition of authorization codes, linked to individual users, poses a challenge for the attackers. Requiring explicit authorization makes it largely impractical to obtain users' authorization codes at a large-scale. 
While attackers may resort to creating malicious \m{}s to deceive users and obtain their authorization codes, replicating this process for other \m{}s is not feasible. Consequently, access tokens obtained through this grant type offer better security against the attacks considered in our study.

\subhead{\M{} to developer server authentication} 
Although our work is mostly centered on \m{} to \s{} authentication using the \a{}s, we also manually test the authentication of \m{} to its developer server. Through experimenting with our own \m{}, we confirm that the developer server does not enforce restrictions for incoming requests from \m{}s. We also confirm that it is possible for a \m{} to communicate with other \m{}s' developer servers. Thus, keeping the secrets in the developer server just shifts the authentication problem if it is not handled by \mbox{the developers properly.}

\subhead{Ways forward} 
We enumerate a few potential directions below. %

\emph{Replacing the developer server with cloud features}.  
As one important advantage of \m{}s, developers can be saved from the need to deploy their own servers. %
They can make use of WeChat cloud base and Tencent cloud hosting to host their
\m{}'s backend, in which no explicit \a{} or access token is required---an implicit grant is implemented. 
In addition, all necessary server-side APIs can be invoked from cloud functions to lessen the reliance on access tokens. However, this is secure only if the \a{} has not been hard-coded, otherwise without IP whitelisting, cloud functions can be called without authorization as seen in Sec.~\ref{sec:callabilityw}.
In certain cases, a developer server is not avoidable, e.g., the business may involve a large amount of data/code, not manageable or cost-effective for the cloud base; in such cases, the developers must avoid hard-coding \a{}s and enable IP whitelisting.

\emph{Mandating IP whitelisting}.
When a developer server is necessary and can be used to make \s{} server-side API calls, IP whitelisting can be enabled to only allow making calls from the IP address of the configured developer server. Despite its technical feasibility, forcing a fixed set of IP addresses may not work for developers who do not own the infrastructure or have servers of a varying nature. But this might be mandated to popular, security-critical \m{}s.

\emph{Disallowing \a{} hard-coding}.
A straightforward way is to restrict it from the source by WeChat/Baidu. As of this writing, WeChat already prevents \m{}s from being released if hard-coded secrets are detected and the latest version of devtools~\cite{weixindevtool} supports \a{} detection as part of code quality analysis. However, this is not retrospective, leaving still 32.6\% live \m{}s with the \a{}s hard-coded. 
We strongly suggest that such hard-coding prohibition should be implemented for all \m{}s. Note that, we even observed from our manual code review, several \m{}s hard-coded valid \a{}s for no apparent reason (i.e., no use of developer/WeChat server-side APIs), and simply add \a{}s in the \texttt{globalData} object. %

\emph{Disallowing server-side API invocation from \m{}s}.
This is already in use through server domain name restrictions~\cite{doc-serverdninfo}, which requires any domain name contacted by a \m{} must be configured in the portal. By disallowing \textit{"api.weixin.qq.com"} (WeChat) and \textit{"openapi.baidu.com"} (Baidu) to be configured, direct server-side API calls can be prevented. It is possible that the \m{}s we have seen with these direct server-side API calls have been developed before this restriction was enforced.

\emph{Switching to \s{} bound dynamic secrets}.
As mentioned earlier, Paytm and Alipay's use of user-bound (as opposed to \m{}-bound in WeChat, Baidu) authorization tokens, generated and managed by the \s{}, replaces fixed \a{}s. While binding to a user involves the user to click, this may be necessary since otherwise the initial trust remains a question, e.g., without prompting the user explicitly, ensuring the request's authenticity is difficult.

\iffalse
\subhead{Safety guidelines by WeChat} 
\V{Advance to the beginning?} \\
WeChat has provided a safety guideline document on sensitive information leaks happening via \m{}s \cite{securitydoc}. The document states that sensitive information such as \a{}, encryption keys, etc should not be encoded in plain text, reversible by Base64 or be present in any other form that appears within the \m{} file. The document also states that if any \m{} is found to have the sensitive information problem, the services to that particular \m{} will be suspended and the \m{} will be removed from the WeChat Open Platform. 
WeChat's explicit statements about security show that they are aware of the issue of hard-coding the \a{} in the \m{}s. Hard-coded credentials have been a problem for a long time, but our research is to uncover how widespread the issue is with \m{}s and other non-compliant practices during development and their security consequences.
As we did our initial automated study of 115,392 WeChat \m{}s, 36,425 \m{}s returned valid access tokens, meaning that these \m{}s are still present on the WeChat open platform. 
\fi

\subhead{Limitations} 
The \m{}s obtained from the WeChat platform whether collected manually or through the use of crawlers, might typically undergo obfuscation, as described in Sec.~\ref{sec:implementation}. The  unpacker utilized in our framework, which is widely recognized within the community, is considered the most popular tool for unpacking \m{}s. However, while utilizing this tool, approximately 4\% of WeChat \m{}s were not unpacked completely, leading to the generation of obfuscated files. This affects the analysis, as our framework performs static analysis on the \m{} code to detect the hard-coded \a{}s. If \a{}s are present within the obfuscated code, our framework may inadvertently overlook them during the analysis. Secondly, as part of our analysis, we installed the WeChat client application on a rooted Android device to access the \m{} packages. Our recent experiment showed that any WeChat account created on a rooted device would be blocked from further use on that device.

 \vspace{-10pt}
\section{Disclosure \& Ethical  Considerations}
We have taken careful consideration to contemplate the ethical implications when designing our analysis framework.
We had to generate the access tokens for the corresponding 
53,377 \m{}s comprising both WeChat and Baidu,
to verify if the identified secrets can lead to security and privacy exposures; note that secrets can be changed by developers, and enabling IP whitelisting can make leaked secrets unexploitable. 
We refrained from calling any Modify API to avoid modification or deletion of \m{} data; we verified such APIs only on our test \m{} we created for research. 
To understand and measure the real and immediate threat to \m{}s with valid access tokens, we call only those Get APIs whose results can be used to call the Modify APIs in order to verify if a \m{}'s data can be modified. We also did not arbitrarily call any server-side API to avoid resource exhaustion. 
According to the regulations from our university's Research Ethics Unit, we took multiple precautions to prevent the exposure of secrets from our collection database.
We did not store any unnecessary data returned from the calls and erased the minimally captured data appropriately after the analysis. 
We also have disclosed our findings to Tencent and Baidu, 
including the list of all the server-side APIs that can be abused when the \a{} is known. Tencent notified us, stating, ``For increments, we will have a review mechanism, and if hard-coded is detected, it will not be approved for release. \textit{For inventory, we will notify developers to fix it, but some developers don't fix it, such is the case in your report}.'' Although we successfully reported to Baidu using their portal, it was more complicated than Tencent.
The email to their security team (security@baidu.com)
was declined because of authentication requirement, and the report required a Chinese phone number to be bound to the reporting account.

\vspace{-5pt}
\section{Related Work}
\label{sec:relwork}
Zhang et al.~\cite{minicrawler} implemented the first large-scale WeChat \m{} crawler and performed an empirical study on the crawled \m{}s. Also, Hao et al.~\cite{hao2018analysis} studied the key features, system architecture, and development prospects of WeChat \m{}s. 
Below we summarize the studies more relevant to our work. 

\subhead{{Security analysis on \m{}s}}
Zhang et al.~\cite{zhang2022small} analyzed the \m{}s permission model of 9 \s{}s, and found six vulnerabilities with at least one security issue in each \s{}; they also presented three proof of concept attacks that can reveal user location, contacts, and clipboard content to unauthorized \m{}s..
Lu et al.~\cite{lu2020demystifying} 
studied security vulnerabilities in 11 \s{}s 
based on the resource management between the \s{}
and the \m{}. Further, Zhang et al.~\cite{zhang2022identity} identified the novel identity confusion based on the app ID, domain name, and capability in 47 high-profile super apps 
based on the identity check adopted by the \s{}s. They demonstrated several attacks based on this vulnerability, such as installing malware on victim's phone, stealing victim's financial accounts, and bypassing security patches.
Furthermore, the National Computer Network Emergency Response Technical Team tested around 50 personal banking \m{}s from WeChat and reported that more than 60\% of the \m{}s did not encrypt the user information both in the device and while it was transmitted~\cite{privacyminibank, privacyminibank1}.
Yang et al.~\cite{yang2022cross} implemented CMRFScanner to identify the cross \m{} request forgery (CMRF) attacks, a novel attack leading to several security consequences, e.g., privileged data access, information leakage, and shopping for free. 
Wang et al.~\cite{wang2023you} developed a consistency analysis framework to identify the inconsistencies between privacy policies and the data practice in the \m{}s. They crawled 10,000 \m{}s from WeChat and extracted 2,998 \m{}s in which they found 2,680 \m{}s did not meet the policy requirements.

Unlike the prior studies on the security of \m{}s,
our study focuses on the \m{}'s \emph{developer server to \s{}} authentication problem, mainly affecting \m{} data that can be user-specific or shared among all users of a \m{}, instead of local resources on a user's phone.
By measuring the extent of insecure development practices (defeating such authentication), we analyze how the resulting unauthorized server-side API calls can cause severe security issues, including bringing down the \m{}s and their services. 
In a concurrent study, Zhang et al.~\cite{zhang2023don} identified 40,880 \m{}s (approximately 1.18\% of the total 3,450,586) that leaked their own \a{}s. However, their study focused on identifying vulnerable \m{}s that leaked their own secrets. We considered both this self-leakage and the leakage of other \m{}s' secrets, resulting in a significantly higher percentage (32.8\%) of \a{} leakage (albeit mostly self-leakage, see the last paragraph in Sec.~\ref{temporalcomp}).

\subhead{Identifying hard-coded secrets} 
Sinha et al.~\cite{sinha2015detecting} provided practical solutions to detect, prevent and fix API key leaks in the source code repositories (GitHub). 
Meli et al.~\cite{meli2019bad} studied the large-scale secrets leakage with GitHub Search API and BigQuery snapshot, for a period of six months,
especially targeting 11 different platforms. CredMinder~\cite{dong2018understanding} is aimed at finding credentials that are leaked in Android apps, by using code analysis instead of string matching, in order to identify credentials even when they are obfuscated. Wen et al.~\cite{wen2018empirical} developed iCredFinder to fix the gap of credential leak detection in iOS apps. Saha et al.~\cite{saha2020secrets} developed a generalized machine learning-based framework with regular expressions to identify the secrets in source code, and analyzed 24 different types of secrets with precision and recall rate of 59\% and 97\% respectively. 
Several other studies focus on OAuth and SSO-related vulnerablities~\cite{wang2015vulnerability,hu2014application, wang2016achilles}; e.g., MoSSOT~\cite{shi2019mossot} 
detects app secret and access token leakage from the network traffic between the relying party and provider apps, which also includes WeChat and its relying third-party apps (but not the \m{}s).

\vspace{-10pt}
\section{Conclusion}
\label{sec:conclusion}
We have presented the \a{} leakage issues in popular \m{} platforms, which is the result of non-compliance of \m{}s with the \s{}'s security guidelines. In order to identify this non-compliance in a large number of \m{}s, we developed a tool to detect the hard-coded \a{}s in the \m{}'s source code and accessibility of the server APIs using the identified secret. We analyzed 110,993 WeChat \m{}s and 10,000 Baidu \m{}s, out of which 36,425 WeChat \m{}s and 112  Baidu \m{}s are found to have valid \a{}s hard-coded. We also discussed how these hard-coded secrets can lead to the misuse of server-side APIs by defeating the authentication mechanisms implemented by the \s{}s. 
Additionally, we used the measurement results to identify potential attack vectors on the identified vulnerable \m{}s that could lead to severe security consequences at the \m{} level.

\vspace{20pt}

\bibliographystyle{abbrv}
\nocite{*}
\bibliography{main}

\appendix
\begin{appendix}
\label{appendix}

\section{Appendix A}
% \am{I suggest deleting this regular expression. First it might have a mistake that can be caught by a reviewer. Second, it shows that the task is too simple}
% \subhead{Regular expression used to identify the app secrets in miniapp's source code}

% [\&]*(SECRET|secret|SEC|sec|Secret|Sec)\textbackslash s*(=|:|([=]*["]*[)]*[.]\newline concat[(])*)\textbackslash s*(")*[a-zA-Z0-9]\{32\}(?!\^{}[0-9]*\$)(?!\^{}[a-zA-Z]*\$)(")*

% \am{Although we are allowed unlimited Appendix, we should keep only the relevant parts. If the reader can understand the paper without any of the material below, do not include it.}

\subsection{WeChat Login} \label{wxlogin}
\begin{figure}[H]
    \includegraphics[width = 0.5\textwidth]{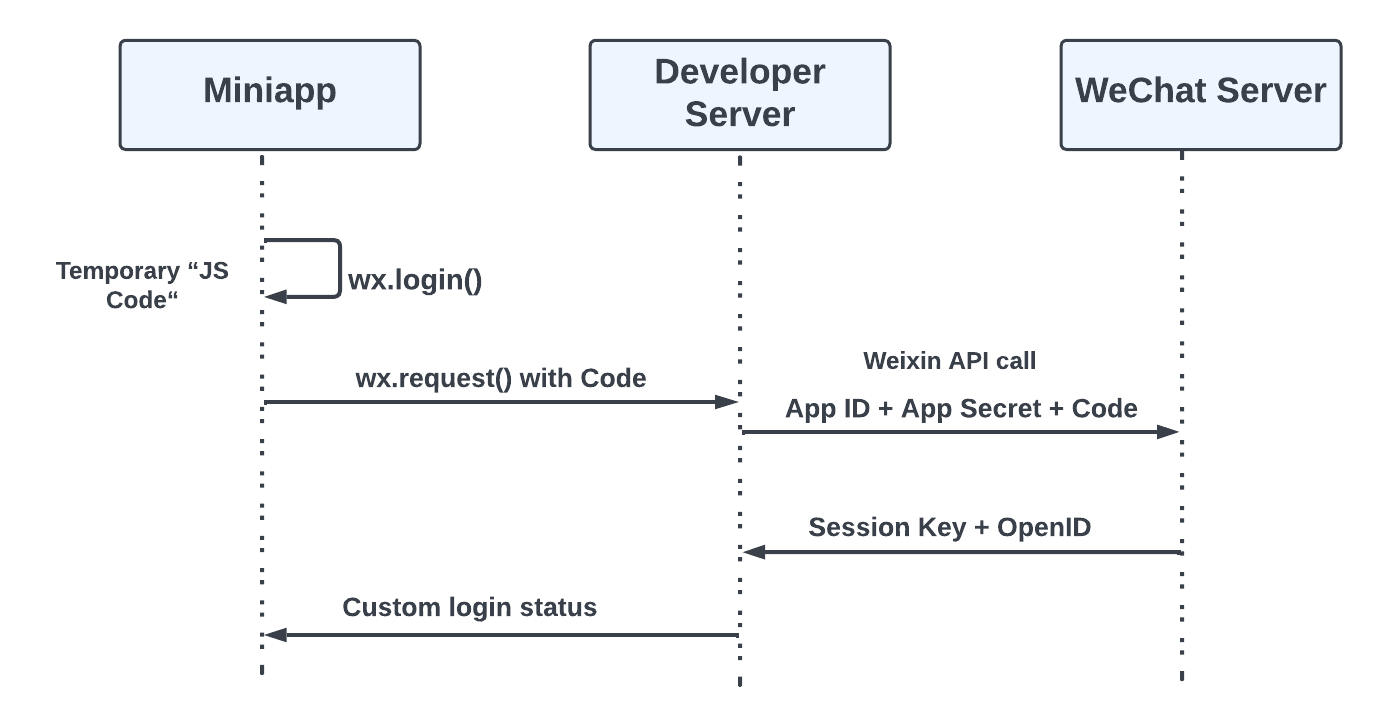}
    \caption{WeChat Code2Session - Login API}
    \label{fig:login}
\end{figure}

\texttt{wx.login} is one of the several APIs provided by WeChat for the developers to interact with the WeChat's native app functions and services (also termed as ``JSAPI''). Whenever the \texttt{wx.login} interface is called in a miniapp, the interface will return a random temporary WeChat login credential (termed as ``JS Code'') which is valid for only 5 minutes. The JS Code is for onetime use only. 
This JS Code is then sent to the developer's server using the interface wx.request, from where the WeChat's Code2Session API \cite{code2session} is called. The WeChat's back-end server returns the session key and Open ID of the miniapp user to the developer server. Figure \ref{fig:login} illustrates the recommended flow of the WeChat Code2Session user authentication API. 
For security reasons, this session key should not be returned to the miniapp as recommended by WeChat \cite{code2session}. The session key will expire only when the miniapp user does not use the corresponding miniapp for a long time. This code to session login interface is to authenticate the miniapp users to the WeChat, and a custom login status is determined to the miniapp.
\label{wechat-login}

\subsection{CVSS Scores} \label{cvssdetails}
% \am{I would move this section to the appendix}
% \supraja{
%We further investigate to measure the effect of each \s{} server-side API call to determine the extent of the detected security consequence. 
We make use of the CVSS (Common Vulnerability Scoring System) score calculator from NVD~\cite{cvsscalc} to calculate the base metrics (confidentiality, integrity and availability), and estimate the security consequence of each \s{} server-side API call. We set our attack vector as \textit{network} and attack complexity as \textit{low}, as calling these APIs does not require any special privileges and involves no user interaction. The scope 
% \mm{} 
of attack does not change in most cases, except for the WeChat cloud base related attacks, where a vulnerable \m{} could affect other non-vulnerable \m{}s or WeChat official accounts; see Sec.~\ref{cloudconsequences}. 
The CVSS scores are summarized in Table~\ref{tab:api-imapct} (and Table~\ref{tab:api-imapct-new} in Appendix). It is to be noted that by the impact of medium and low, it defines the impact of the attack on the \m{} and the user. This does not change the attacker capabilities and hence, for executing all these \s{} server-side APIs, the attacker requires only a network connection.

\subsection{Unauthorized Invocation of Baidu Server-side APIs}
\label{sec:callabilityb}
\subhead{\\Message templates} We test this category of APIs to verify if the message templates of the \m{}s can be accessed by an attacker.
% \mm{ to do what}. 
We see from our analysis that the \texttt{getTemplateLibraryList} API returns valid values for all 112 (100\%) \m{}s, and hence it is possible 
% \mm{how do you know?} 
to call the \texttt{addTemplate} API for these 112 \m{}s, as the output of the former API can be used as an input for the latter. Using our framework, we call \texttt{getTemplateList} API for all 112 \m{}s, which returned valid values for 74 (66\%) \m{}s, and thus it is possible to call \texttt{deleteTemplate} for these 74 \m{}s.

\subhead{Traffic distribution resources}
Using this category of APIs, a \m{} developer can distribute the resources for a \m{} across different paths. We test this list of APIs to check if an attacker is capable of submitting resources such as image files to the Baidu server.
% \am{What do you mean by materials} on behalf of the \m{} 
%maliciously. 
We did not actually make calls using these APIs (\texttt{submitResource}, \texttt{submitSitemap}, \texttt{interfaceSubmission}, \texttt{submitsku}), but verified only the callability. These calls require several parameters, which are all user-controlled values, and thus the APIs can be called by the attacker for all the 112 (100\%) \m{}s with valid \a{}s.

\subhead{Coupons}
We test the APIs under this category to verify if an attacker can create and manage Baidu coupons (provided by Baidu to the \m{} developers, who then  manage and distribute the coupons to the \m{} users).
% \mm{=?}
We do not call any APIs under this category to avoid adding anything to the server arbitrarily, and thus only check the callability of these APIs (\texttt{createCoupon}, \texttt{submitCoupon}, and \texttt{ManageCoupon}). For all 112 (100\%) \m{}s with valid \a{}s, these APIs are callable, hence enabling an attacker to manipulate the coupons.
\begin{figure}[H]
    \includegraphics[width = 0.5\textwidth]{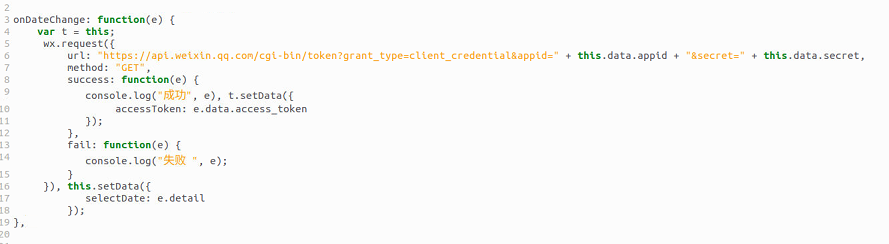}
    \caption{Example code snippet of a WeChat \m{} using server-side API calls in the code with hard-coded secret}
    \label{fig:code-snippet1}
\end{figure}
\begin{figure}[H]
    \includegraphics[width = 0.5\textwidth]{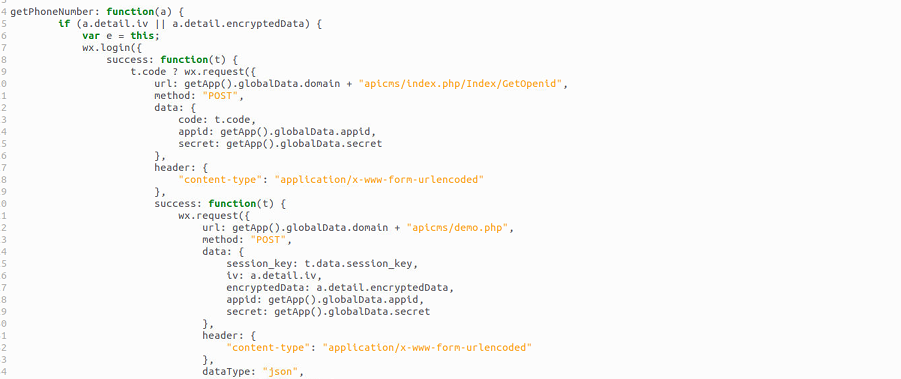}
    \caption{An example of a WeChat \m{} calling a developer server API for WeChat login with a hard-coded secret that is passed from the \m{}.}
    \label{fig:code-snippet-api2}
\end{figure}
\begin{table}[H]
\centering
\begin{tabular}{l | l }
\toprule
\textbf{API Categories} & \textbf{\# miniapps} \\
\midrule
\M{} User Login & 1723 (42\%)\\
Access Token &  554 (13.5\%)  \\
Customer Service & 429 (10.4\%) \\
Dynamic Messages & 319 (7.7\%)\\
Generate QR code & 285 (6.95\%) \\
Message Templates & 279 (6.8\%) \\
Image Processing & 176 (4.3\%) \\
Security Check & 134 (3\%) \\
Live Broadcast & 81 (1.97\%) \\
Logistics & 77 (1.87\%) \\
Data Analytics & 41 (1\%) \\
\bottomrule
\end{tabular}
\captionsetup{skip=2pt}
\caption{WeChat \m{}s with the direct invocation of server-side APIs} %
% \mm{my understanding is that all these calls will fail. In that case, should we have this table or not?} \V{I remember Supraja was not able to find out... am I right?} \supraja{Supraja: Theroetically, yes these calls should fail. Practically, I tested this with our own miniapp and it was not possible to even add the domain (expected). But still unsure if these would work or fail in the real cases.}\mm{then you should remove it or put it in the appendix}
\label{tab:apidirect}
\end{table} 
\begin{table}[htb]
\centering
\begin{tabular}{l|rrr}
\toprule
\textbf{Consequences} 
& \textbf{\shortstack{DATASET1}}& \textbf{\shortstack{DATASET3}}&
\textbf{\shortstack{DATASET2}}\\
\midrule
% \rowcolor{gray(x11gray)}
\# read \m{} data &  36,425 & 2,572 &112\\
\# send msgs. & 18,241 & 199& -\\
% \rowcolor{gray(x11gray)}
\# data tampering & 36,425 &2,572  &112\\
%\shortstack{\#  of \m{}s\\with hard-coded\\\a{}s}
\# malicious redirects
& 23,162 &573   &- \\
% \rowcolor{gray(x11gray)}
%\shortstack{\#  of generated\\access tokens} 
\# resource exhaustion 
& 36,425 & 2,572 &112 \\
\bottomrule
\end{tabular}
\vspace{-10pt}
    \caption{Relationship between number of WeChat (DATASET 1 and DATASET 3) and Baidu (DATASET 2) \m{}s against the identified security consequences. 
    \newline
    \newline}%
    \vspace{-15pt}
    \label{tab:consequences}
\end{table}
\begin{table}[H]
\centering
\resizebox{\linewidth}{!}{
\begin{tabular}{l l l}
% \multirow{}{*}[1ex]{\textbf{API Category}} & 
% \multirow{}{*}[1ex]{\textbf{Get API}} & 
% \multirow{}{*}[1ex]{\textbf{Modify API}}&  
\textbf{API Category} & 
\textbf{Get API} & 
\textbf{Modify API}\\  \hline
\multirow{2}{*}{Access Token} &getAccessToken &-\\[10pt]
\hline
\multirow{3}{*}{Customer Service Message} & -&uploadTempMedia\\[4pt]
 &  &customerServiceMessage.send\\[3pt]
\hline
% \multirow{4}{*}{Message Templates} &getPubTemplateTitleList &addTemplate\\[3pt]
%  & getMessageTemplateList &deleteMessageTemplate\\[3pt]
%  & getCategory  &sendTemplateMessage\\[3pt]
% \hline
\multirow{2}{*}{Updatable Message} &createActivityId &setUpdatableMsg\\[10pt]
\hline
\multirow{3}{*}{Miniapp Plug-in} & &managePlugin\\[3pt]
& &managePluginApplication\\[3pt]
\hline
% \multirow{2}{*}{Uniform Messages} &- &sendMessage\\[10pt]
% \hline
\multirow{2}{*}{Miniapps Nearby} &getNearbyPoiList &deleteNearbyPoi \\[3pt]
 & &setShowStatus\\[3pt]
\hline
% \multirow{4}{*}{Miniapps Code} &createQRCode &- \\[3pt]
% &wxacode.get &\\[3pt]
% &getUnlimited \\[3pt]
% &generateShortLink \\
% \hline
\multirow{2}{*}{Logistics Assistant} &getPrinter &updatePrinter\\[3pt]
% &getBindAccount &bindLocalAccount \\ [3pt]
% &getAllAccount & \\ [3pt]
&getAllDelivery &\\[3pt]
\hline
% \multirow{2}{*}{Transaction} &getGuaranteeStatus & \\
% &getPenaltyList & \\ [3pt]
% \hline
\multirow{2}{*}{openAPI Management} &getApiQuota &clearQuotaByAppSecret \\
& &clearQuota \\ [3pt]
\hline
\multirow{2}{*}{Operations and Maintenance} & getFeedback & \\
&getDomainInfo & \\ [3pt] \hline
% &getPerformance & \\ [3pt] \hline
\multirow{5}{*}{Cloud Base} & &invokeCloudFunctions\\
&databaseCollectionGet &databaseCollectionAdd\\
& &databaseCollectionDelete\\
&getQcloudToken &databaseAdd \\
& &databaseDelete \\
&
 \                     &databaseUpdate \\
& &databaseQuery \\ \hline
% \multirow{3}{*}{Data Analytics} &get all Monthly visit Data &\\
% &get all weekly visit data &\\
% &get all daily data &\\
% \hline
\end{tabular}}
    \caption{List of get and modify WeChat server-side APIs evaluated in our measurement study chosen based on the availability of request parameters.} %
    \label{api-list-tested}
\end{table} 
\begin{table}[H]
\centering
\resizebox{\linewidth}{!}{
\begin{tabular}{l l l}
\textbf{API Category} & 
\textbf{Get API} & 
\textbf{Modify API}\\
\hline
\multirow{2}{*}{Access Token} &getAccessToken &-\\[10pt]
\hline
% \multirow{1}{*}{Customer Service Message} & - &customerServiceMessage.send\\[3pt]
% \hline
\multirow{4}{*}{Message Templates} & &addTemplate\\[3pt] 
 & getTemplateList &deleteMessageTemplate\\[3pt]
\hline
% \multirow{2}{*}{Subscribe news} & - &subscribeSend\\[10pt]
% \hline
\multirow{2}{*}{Traffic Distribution Resources} &  &submitResource \\[3pt]
 & - &submitSitemap \\[3pt]
& &interfaceSubmission \\[3pt]
& &submitsku \\[3pt]
\hline
\multirow{2}{*}{Coupons} &  &createCoupon \\[3pt]
 & - &submitcoupon \\[3pt]
& &ManageCoupon \\[3pt]
\hline
\end{tabular}}
    \caption{List of get and modify Baidu server-side APIs evaluated in our measurement study chosen based on the availability of request parameters.} %
    \label{baidu-api-list-tested}
\end{table} 
\begin{table*}[t]
\centering
\resizebox{\linewidth}{!}{
\begin{tabular}{l|llrccccl}
\textbf{WeChat server-side APIs}&
\textbf{Required parameters} &
% \textbf{Invocation} &
\textbf{\# miniapps}&
% \begin{rotate}{60}\textbf{Arbitrary Uploads} \end{rotate}&
\textbf{[A]}&
\textbf{[B]} &
\textbf{[C]} &
% \begin{rotate}{60}\textbf{ DB Compromise} \end{rotate} &
\textbf{[D]} & 
\textbf{[E]} & 
\textbf{Impact}\\ 
\hline
\textbf{clearQuotaByAppSecret}          & appID, appSecret  &2,572    &   &   &\textcolor{red}\checkmark   &  &\textcolor{red}\checkmark & High \\
% \hline
\textbf{clearQuota}          & AT, appID &2,572   &   &   &\textcolor{red}\checkmark  & &\textcolor{red}\checkmark  & High \\ 
% \hline
\textbf{managePlugin}            & AT, pluginAppID  &433   &\textcolor{red}\checkmark   &   &\textcolor{red}\checkmark    & &\textcolor{red}\checkmark  & High \\ 
% \hline
\textbf{deleteNearbyPoi}               & AT, poiID  &303   &   &   &\textcolor{red}\checkmark    &  &\textcolor{red}\checkmark & High \\ 
% \hline
\textbf{setShowStatus}           & AT, poiID  &303   &   &   & \textcolor{red}\checkmark   &  &\textcolor{red}\checkmark & High \\ 
% \hline
\textbf{managePluginApplication} & AT, appID   &179  &\textcolor{red}\checkmark   &   &\textcolor{red}\checkmark    &   &\textcolor{red}\checkmark & High\\ 
% \hline
\textbf{invokeCloudFunctions}   & AT, CloudFunctionName  &65  &\textcolor{red}\checkmark   & \textcolor{red}\checkmark  &\textcolor{red}\checkmark  &   & \textcolor{red}\checkmark  & High \\ 
% \hline
\textbf{databaseCollectionGet}   & AT, CloudEnv   &21   & \textcolor{red}\checkmark  &   & & &\textcolor{red}\checkmark & High \\ 
% \hline
\textbf{databaseCollectionAdd}   & AT, CloudEnv, CollectionName   &21  &   &   &\textcolor{red}\checkmark    & &\textcolor{red}\checkmark & High \\ 
% \hline
\textbf{databaseCollectionDelete}&AT, CloudEnv, CollectionName   &21   &   &   &\textcolor{red}\checkmark  &&\textcolor{red}\checkmark & High \\ 
% \hline
\textbf{databaseAdd}             & AT, CloudEnv   &21   &   &   &\textcolor{red}\checkmark   & &\textcolor{red}\checkmark & High \\ 
% \hline
\textbf{databaseDelete}          & AT, CloudEnv  &21 &   &   &\textcolor{red}\checkmark   & &\textcolor{red}\checkmark & High \\ 
% \hline
\textbf{databaseUpdate}          & AT, CloudEnv   &21 &   &   &\textcolor{red}\checkmark   & &\textcolor{red}\checkmark & High \\ 
% \hline
\textbf{databaseQuery}           & AT, CloudEnv   &21  &\textcolor{red}\checkmark   &   &\textcolor{red}\checkmark   & & \textcolor{red}\checkmark& High \\ 
% \hline
\textbf{setUpdatableMsg}         & AT   &4   &   &\textcolor{red}\checkmark    &\textcolor{red}\checkmark  &\textcolor{red}\checkmark & \textcolor{red}\checkmark  & High\\ 
% \hline

\textbf{uploadTempMedia}         & AT  &2,572   &    &   & \textcolor{red}\checkmark  &  &\textcolor{red}\checkmark &Medium \\ 
% \hline
\textbf{getApiQuota}          & AT, cgi\_path   &2,572   &\textcolor{red}\checkmark   &   &    & & \textcolor{red}\checkmark & Medium \\ 
% \hline
\textbf{getDomainInfo}          & AT   &2,101  &\textcolor{red}\checkmark   &   &   & &\textcolor{red}\checkmark  & Medium \\ 
% \hline
\textbf{getAllDelivery}              & AT  &252   & \textcolor{red}\checkmark  &     &   & &\textcolor{red}\checkmark & Medium \\ 
\textbf{customerServiceMessage.send}  & AT, openID    & 151 &   & \textcolor{red}\checkmark  &  &\textcolor{red}\checkmark  &\textcolor{red}\checkmark & Medium\\ 
% \hline
\textbf{getPrinter}              & AT   &114 &\textcolor{red}\checkmark   &   &  &  &\textcolor{red}\checkmark & Medium \\ 
% \hline
\textbf{updatePrinter}           & AT, openID    &114 &   &   &\textcolor{red}\checkmark   & & \textcolor{red}\checkmark & Medium \\ 
% \hline
\textbf{getFeedback}          & AT   &93   &\textcolor{red}\checkmark   &   &   & &\textcolor{red}\checkmark  & Medium \\ 
% \hline
\textbf{getQcloudToken}          & AT   &59  & \textcolor{red}\checkmark  &   &   & &\textcolor{red}\checkmark  & Medium \\ 
% \hline

\textbf{createActivityId}        & AT   &2,572  & \textcolor{red}\checkmark   &   &   &  &\textcolor{red}\checkmark & Low \\ 
% \hline
\textbf{getNearbyPoiList}         & AT   &303   &\textcolor{red}\checkmark   &   &   & &\textcolor{red}\checkmark  & Low \\ 
\hline
\end{tabular}}
\caption{Statistics of unauthorized callable WeChat  server-side APIs for \texttt{DATASET3}. 
% Since \texttt{invokeFunctions} calls  functions are implemented by developers, the actual impact vary a lot depending on what the function does \am{Why this is needed in the caption?}. 
AT: Access Token; [A]: Read \M{} Data; [B]: Send Arbitrary Messages; [C]: Data Tampering; [D]: Malicious Redirects; [E]: Resource Exhaustion; \textcolor{red}\checkmark\  denotes the possibility of the attack using the corresponding -side API.
}%
\label{tab:api-imapct-new}
\end{table*}
\end{appendix}

\end{document}